\documentclass[journal,twocolumn,10pt]{IEEEtran}
\IEEEoverridecommandlockouts 
\usepackage{epsfig,latexsym}
\usepackage{float}
\usepackage{indentfirst}
\usepackage{amsmath}
\usepackage{amssymb}
\usepackage{times}
\usepackage{subfigure}
\usepackage{pifont}
\usepackage{psfrag}
\usepackage{cite}
\usepackage{url}
\usepackage{lastpage}
\usepackage{bm}
\usepackage{color}
\usepackage{fancyhdr}
\usepackage{balance}
\usepackage{fancyhdr,lastpage}
\usepackage{epstopdf}
\begin{document}
\title{A Unified Framework for IRS Enabled Wireless Powered Sensor Networks}
\author{\IEEEauthorblockN{Zheng Chu,
	 \emph{Member, IEEE}, 
Zhengyu Zhu,
 \emph{Member, IEEE}, 
Miao Zhang,
 \emph{Member, IEEE}, 
 Fuhui Zhou,
 \emph{Senior Member, IEEE},
Li Zhen,
 \emph{Member, IEEE}, 
 Xueqian Fu, \emph{Member, IEEE},
 and
 Naofal Al-Dhahir, \emph{Fellow, IEEE}
}
}
\vspace{-0.2in}
\maketitle 
\thispagestyle{empty}
\vspace{-0.3in}
\begin{abstract}
This paper unveils the importance of intelligent reflecting surface (IRS) in a wireless powered sensor network (WPSN). 
Specifically, a multi-antenna power station (PS) employs energy beamforming to provide wireless charging for multiple Internet of Things (IoT) devices, which utilize the harvested energy to deliver their own messages to an access point (AP). 
Meanwhile, an IRS is deployed to enhance the performances of wireless energy transfer (WET) and wireless information transfer (WIT) by intelligently adjusting the phase shift of each reflecting element. 
To evaluate the performance of this IRS assisted WPSN, we are interested in maximizing its system sum throughput to jointly optimize the energy beamforming of the PS, the transmission time allocation, as well as the phase shifts of the WET and WIT phases. The formulated problem is not jointly convex due to the multiple coupled variables. To deal with its non-convexity, we first independently find the phase shifts of the WIT phase in closed-form. We further propose an alternating optimization (AO) algorithm to iteratively solve the sum throughput maximization problem. To be specific, a semidefinite programming (SDP) relaxation approach is adopted to design the energy beamforming and the time allocation for given phase shifts of WET phase, which is then optimized for given energy beamforming and time allocation.
Moreover, we propose an AO low-complexity scheme to significantly reduce the computational complexity incurred by the SDP relaxation, where the optimal closed-form energy beamforming, time allocation, and phase shifts of the WET phase are derived.
Finally, numerical results are demonstrated to validate the effectiveness of the proposed algorithm, and highlight the beneficial role of the IRS in comparison to the benchmark schemes.
\end{abstract}
\IEEEpeerreviewmaketitle
\setlength{\baselineskip}{1\baselineskip}
\newtheorem{definition}{Definition}
\newtheorem{fact}{Fact}
\newtheorem{assumption}{Assumption}
\newtheorem{theorem}{Theorem}
\newtheorem{lemma}{Lemma}
\newtheorem{corollary}{Corollary}
\newtheorem{proposition}{Proposition}
\newtheorem{example}{Example}
\newtheorem{remark}{Remark}
\newtheorem{algorithm}{Algorithm}
\section{Introduction}
In recent years, demands for massive connectivity and high data rate have experienced an explosive growth in next generation wireless networks, which has drawn increasing attention of academics and industries. Internet of things (IoT) is considered as an important portion of the fifth-generation (5G) network and beyond, which significantly enhances high-data access rate for massive wireless devices (WDs) \cite{Dobre_WC_2018,Dobre_IMM_2019}. In a generic IoT system, a number of sensor nodes connect with an access point (AP) forming a wireless sensor network (WSN), which has been widely applied in various practical scenarios, including external environment monitoring, event detection for emergency services, wireless surveillance for public safety, healthcare diagnosis, etc. \cite{Zheng_WPSN_IoT_2018}.
The sensors are typically low energy consumption devices, which depend on a
finite-capacity battery such that they may not have sufficient energy to support their own computation and communication operations \cite{Rui_Zhang_WPCN_CM_2015}. The sensors are generally equipped with traditional batteries, which limits their potential practical applications, since traditional batteries need to be regularly maintained or even replaced to prolong their own operation lifetime, which is often costly due to the massive number of sensors in use. In addition, the sensors are often deployed in extreme environments, infrastructures, or human bodies such that the battery maintenance or replacement is very challenging. Thus, limited battery lifetime of sensors is still a key challenge in designing future wireless networks \cite{Rui_Zhang_WPCN_CM_2016}. 

Radio frequency (RF) wireless energy transfer (WET), as one of the promising solutions to address the energy constraint issue, enables novel electric energy transmission from a dedicated energy source to the WDs without any wired connection \cite{Rui_Zhang_TWC_2013}. As one important solution of RF WET, wireless powered communication networks (WPCNs) collect energy from dedicated energy sources, each of which has a stable energy supply, supporting wireless information transfer (WIT). A WPCN is expected to improve the system throughput in comparison to the conventional battery-powered communications. In WPCNs, a classic protocol named
``\emph{harvest-then-transmit}'' was proposed in \cite{RZhang_WPCN_TWC_2014}, where the WDs first collect energy from the RF signals broadcast by an AP, and then transmit their independent information to the AP utilizing the harvested energy. A group of power stations (PSs) constitute of a dedicated WET network which is deployed to support WIT in the vicinity \cite{Kaibin_Huang_CM_2015,Kaibin_Huang_TWC_2014_WPT}. The WDs can gain energy benefits from these PSs to prolong their own battery life via wireless charging. The WPCN improves the energy efficiency in wireless networks which reduces its operational cost. 
Thus, the WPCN is more suitable for the low energy consumption use cases, i.e., wireless powered sensor networks (WPSNs), which includes many sensor nodes to reduce maintenance cost as well as enhance its deployment flexibility. In addition, the WPSN is generally used to support low-power devices, i.e., radio frequency identification (RFID) IoT sensor nodes and tags \cite{Rui_Zhang_WPCN_CM_2016}.
 
 Moreover, 5G and beyond networks have been evolving towards machine-centric driven by a vast range of quality of service (QoS) requirements, e.g., ultra-high spectral efficiency and throughput. 
  Several advanced techniques can be used to enhance spectral efficiency and throughput, e.g., massive multiple-input multiple-output (massive MIMO) with millimetre wave (mmWave), relaying, and ultra-dense networks (UDNs) \cite{ZGao_WC_2015,XChen_CM_2015,SChen_WC_2016}. 
  However, these techniques typically require a large amount of RF chains over a high frequency band and incur high energy consumption and hardware cost, which results in more energy consumption \cite{RMarcoDi_EURASIP_IRSTUT_2019,Renzo_JSAC_2020}. This has given rise to a novel and promising paradigm, named \emph{smart radio environment}, which is a holographic wireless mode with low cost, size, weight, and power consumption features in hardware architecture \cite{CHuang_WCM_2020}. \emph{Smart radio environment} offers the seamless wireless connectivity and the capability of transmitting and processing data via recycling the existing radio waves instead of generating new ones, which is a transformative way to convert the traditional wireless environment into a programmable intelligent entity \cite{RMarcoDi_EURASIP_IRSTUT_2019,Renzo_JSAC_2020,CHuang_WCM_2020}.

Intelligent reflecting surface (IRS), as an enabler of \emph{smart radio environments}, is one of the promising techniques for enhancing the throughput in future sixth generation (6G) wireless communication networks \cite{SDang_Nature_Ele_2020}. The IRS is made of a massive number of reconfigurable reflecting elements, and coordinated by a software-oriented IRS controller \cite{RMarcoDi_EURASIP_IRSTUT_2019}. These reflecting elements typically have small, low-cost, and low-energy consumption features, which can efficiently reflect the intended signal without a dedicated RF processing, en/de-coding, or re-transmission \cite{QWu_TWC_2019,QWu_CM_2019}. 
The IRS exploits a novel wireless communication paradigm to achieve three-dimensional (3D) passive beamforming gains to align the reflected signals with the direct signal by intelligently varying the phase shift at each reflecting element. 
\subsection{State of the Art}
There are several existing studies in the literature that focus on the IRS assisted wireless communication networks, such as \cite{QWu_TWC_2019,QWu_CM_2019,CHuang_TWC_EEIRS_2019,RZhang_TCOM_2020,ZCHU_WCL_IRS_2019,HShen_IRS_Sec_CL_2019,CPAN_MCELL_2019}.
In \cite{QWu_TWC_2019,QWu_CM_2019}, an IRS assisted multiple-input single-output (MISO) system is proposed, where the desired signal is passively received and reflected by the IRS to align the reflecting link with the direct link between the access point (AP) and the users via intelligently adjusting the phase shift of each reflecting element. The total transmit power was minimized subject to the individual signal-to-interference-plus-noise ratio (SINR) constraint to jointly design the active transmit and passive reflecting beamforming.
The energy efficiency was adopted as a performance metric in the IRS aided MISO downlink system, which is maximized such that the power allocation and the phase shifts of the IRS are optimally designed by an alternating optimization (AO) algorithm, gradient descent search, and sequential fractional programming \cite{CHuang_TWC_EEIRS_2019}. In \cite{RZhang_TCOM_2020}, a practical phase shift model was proposed to capture the phase-dependent amplitude variation in the element-wise reflection design.
Furthermore, the potential of the IRS enabled wireless security was demonstrated in \cite{ZCHU_WCL_IRS_2019,HShen_IRS_Sec_CL_2019}. These works focus on the IRS-assisted MISO secure system, where an IRS is deployed to reduce power consumption and enhance achievable secrecy performance via alternately designing the secure transmit and reflecting beamforming vectors. 
%
In \cite{CPAN_MCELL_2019}, a multi-cell multiple-input multiple-output (MIMO) system was investigated with the assistance of an IRS in enhancing the downlink transmission for cell-edge users to mitigate the inter-cell interference. 
Moreover, the IRS enabled simultaneous wireless information and power transfer (SWIPT) was investigated in \cite{QWu_WCL_2020,QWu_2019,CPAN_IRSSWIPTMIMO}. In \cite{QWu_WCL_2020}, an IRS was introduced in a SWIPT system to achieve high passive beamforming gains. This enhances the WET efficiency and the rate-energy trade-off of the IRS enabled SWIPT system via maximizing weighted harvested energy of energy harvesting users subject to the individual SINR constraint for information decoding users. Extension to the multiple IRSs scenario, which are deployed to support the information/energy transfer from a AP to information decoding/energy harvesting users \cite{QWu_2019}. The total transmit power is minimized with the QoS constraints for all users, which demonstrates the impact of these IRSs on energy efficiency of the SWIPT system. 
In \cite{CPAN_IRSSWIPTMIMO}, an IRS assisted MIMO SWIPT system was studied to maximize the weighted sum rate (WSR) of the information decoding users, while guaranteeing the energy harvesting requirement of the energy harvesting users. The classic block coordinate descent (BCD) algorithm was applied to decompose the formulated problem into several sub-problems and alternately design the transmit precoding and phase shift matrices.

The above-mentioned works focus on the IRS aided downlink transmission. On the other hand, the IRS aided multiple access channel (MAC) has drawn increased research attention recently, such as the IRS enabled MAC \cite{RZhang_MAC_IRS_2020}, the IRS assisted WPCN user cooperation (UC) system \cite{SZBi_IRS_WCL_2020}.
In \cite{RZhang_MAC_IRS_2020}, the capacity region and deployment mechanism of the IRS enabled MAC were investigated for a two-user case with the help of distributed and centralized IRSs. Additionally, an IRS-assisted WPCN with UC was considered in \cite{SZBi_IRS_WCL_2020}, where the IRS is deployed to help a WPCN aided UC system to enhance the common throughput performance in downlink WET and uplink WIT phases. Very recently, the IRS was introduced in mobile edge computing (MEC) systems which coordinates the computation and communication capabilities of wireless networks via local computing on the WDs or offloading to the MEC server for processing their own computational tasks \cite{Bai2019LatencyMEC,Bai2020MEC}.
In the IRS assisted MEC system, the WDs offload a portion of their computational tasks to the AP equipped with a MEC server with the assistance of an IRS \cite{Bai2019LatencyMEC}.
The computational latency is minimized
 to satisfy the constraints of the edge computing capability, which alternately design the computing and communication configurations. In \cite{Bai2020MEC}, a further extension was investigated in terms of the IRS enabled WPCN MEC network, where the IRS assists the WDs via a reflecting link for WET and computational offloading to reduce the total energy consumption of the WET at the AP and of the edge computing based on orthogonal frequency-division multiplexing (OFDM) configurations. 

Although the existing state-of-the-art made variety of research contributions in the downlink/uplink IRS aided wireless systems, the IRS aided SWIPT systems and the IRS assisted MEC systems, there still exists a major research gap on investigation of the IRS's benefits in a WPSN. We introduce the IRS to enhance energy harvesting and data transmission capabilities of the WPSN due to its self-sustainability, which is a promising paradigm and has not yet been investigated in the literature. Also, the IRS is deployed to properly coordinate the energy/information RF signals which can circumvent the high RF signal attenuation due to long distances or construct an effective energy harvesting/charging zone. This provides an coverage enhancement for downlink wireless charging and uplink connectivity so as to improve the throughput performance for the WPSN. In addition, it is imperative that the energy source is equipped with multiple antennas which realizes energy beamforming to perform WET. This is due to the fact that the energy beamforming can bring a significant WET efficiency in comparison to the traditional omnidirectional multi-antenna transmission.
To the best of the authors' knowledge, there has been few of work that modelled and investigated the IRS assisted WPSN with multi-antenna energy source, which motivates our work in this paper.
 
This paper investigates an IRS assisted WPSN, where a multi-antenna PS applies energy beamforming to radiate wireless energy to multiple IoT devices which then utilize the harvested energy to transmit their own information to the AP. The IRS is deployed to improve energy harvesting and data transmission capabilities with intelligently adjustable phase shifts. 
	\emph{For this novel system model, the main contributions of this paper are summarized in the following.}
	\begin{enumerate}
		\item We first investigate an IRS aided WPSN system, where the IRS acts as a helper to enhance the throughput performance of the WPSN by enabling energy and information reflections during the WET downlink and WIT uplink transmissions, respectively.
		\item We maximize the system sum throughput to evaluate the energy harvesting and data transmission capabilities of the IRS assisted WPSN, which jointly designs the energy beamforming of the PS, the passive phase shifts as well as the transmission time allocation of the WET and WIT phases. The formulated problem is not jointly convex, due to multiple coupled variables, and cannot be solved directly. To tackle the non-convexity, we first derive the optimal closed-form phase shifts of the WIT phase. Next, we take into consideration an AO algorithm to alternatively update the energy beamforming, the phase shifts of the WET phase and the transmission time allocation. Specifically, a semi-definite programming (SDP) relaxation scheme is applied to design the energy beamforming as well as the time allocation for given phase shifts of the WET phase which is also solved by the SDP relaxation for given energy beamforming and time allocation. In addition, the rank-profile is characterized to obtain an optimal energy beamforming, while Gaussian randomization is adopted to deal with possible higher-rank phase shift matrix of the WET phase incurred by the SDP relaxation.
		\item Although the SDP based AO scheme can efficiently solve the formulated problem, it incurs a higher computational complexity and is time-consuming, especially for a large amount of reflecting elements. Thus, it is imperative to propose an independent scheme which can be deployed on the IRS and the IoT devices, and significantly lowers the computational complexity incurred by the SDP relaxation with Gaussian randomization. In light of this, the low complexity based AO scheme is proposed. Particularly, the optimal closed-form energy beamforming and transmission time allocation are derived by taking into consideration eigen-decomposition, as well as the Lagrange dual method and the KKT conditions, respectively, for given phase shifts of the WET phase. Then, we derive the closed-form phase shifts of the WET phase by the Majorization-Minimization (MM) algorithm for given energy beamforming and transmission time allocation. In addition, the convergence of the proposed MM algorithm and low complexity based AO scheme are analyzed. 
\end{enumerate} 
The remainder of this paper is organized as follows. Section \ref{section system model} introduces the system model as well as the sum throughput maximization problem formulation. Section \ref{section:STM_IRS_WPSN} provides the optimal solution of the formulated problem. Numerical results are demonstrated in Section \ref{section:simulation} to evaluate the proposed algorithm. Finally, we conclude this paper in Section \ref{section:Conclusion}.
\subsection{Notations}
We use the upper case boldface letters for matrices and lower case boldface letters for vectors. 
$ \textrm{conj}(\cdot) $, $ (\cdot)^{T} $ and $ (\cdot)^{H} $ denote the conjugate, the transpose and conjugate transpose operations, respectively. $ \textrm{Tr}(\cdot) $ stands for trace of a matrix. $ \rho_{max}(\cdot) $ represents the maximum eigenvalue, whereas $ \bm{\nu}_{\max} (\cdot) $ denotes the eigenvector associated with the maximum eigenvalue. $ \mathbf{A} \succeq  \mathbf{0} $ indicates that $ \mathbf{A} $ is a positive semidefinite matrix. $| \cdot |$ and $ \| \cdot \| $ denote the absolute value and  the Euclidean norm of a vector. $ \mathbf{I} $ denotes the identity matrix with appropriate size. $ \exp(\cdot) $ and $  \arg $ indicate the exponential function and the phase angle, respectively. $ \mathcal{W}(\cdot) $ denotes the Lambert $ \mathcal{W} $ function. $ \Re \{ \cdot \} $ represents the real part of a complex number. $\textrm{diag}\{ \cdot \}$ denotes a vector that consists of the diagonal elements of a matrix or a diagonal matrix where the diagonal elements are from a vector.
\section{System Model}\label{section system model}
\begin{figure}[!htbp]
	\centering
	\includegraphics[scale = 0.4]{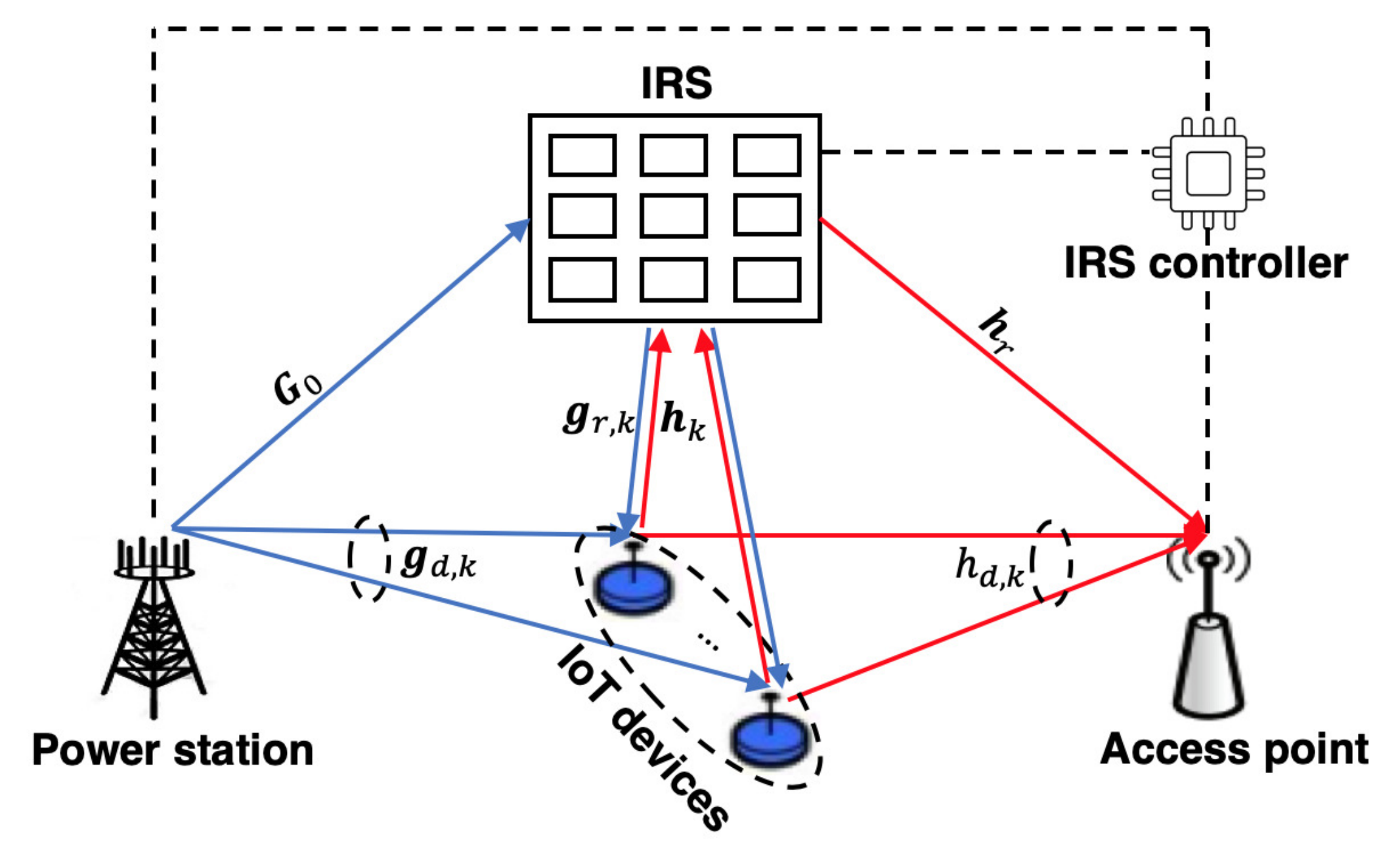}
	\caption{An IRS assisted WPSN.}
	\label{fig:System_model}
\end{figure}
In this paper, we consider an IRS assisted WPSN as shown in Fig. \ref{fig:System_model}. To be more specific, a multi-antenna PS employs energy beamforming to broadcast wireless energy to $ K $ IoT devices who send their own information to an AP using the harvested energy. Meanwhile, an IRS is deployed to enhance energy harvesting and data transmission capabilities by generating passive energy and information passive beamformers. It is assumed that the PS is equipped with $ N_{T} $ transmit antennas, the IRS consists of $ N_{R} $ reflecting elements, while the other devices (i.e., the AP, and all IoT devices) are single-antenna nodes. The IRS controller is generally fixed to coordinate its reflection mode
by intelligently varying the phase shift of each reflecting element.
In this paper, all channel state information (CSI) is assumed to be perfectly known because we are interested in an upper bound on the sum throughput performance in the IRS assisted WPSN \cite{RZhang_WPCN_TWC_2014,QWu_TWC_2019}. 
Several channel estimation techniques have been investigated in the literature for obtaining the CSI of the direct link between the PS and the devices as well as the devices and the AP \cite{RZhang_WPCN_TWC_2014,QWu_WPCN_TWC_2016}.
	Moreover, the IRS is equipped with a controller that coordinates its switching between two working modes, such as reception mode for channel estimation and reflection mode for energy/data transmission \cite{QWu_TWC_2019}, which also provides a real-time CSI feedback via this controller.
	Very recently, a novel channel estimation framework was proposed based on the PARAllel FACtor (PARAFAC) decomposition to unfold the resulting cascaded channel model \cite{CHUANG_C2020,CHUANG_J2020}. 
On the other hand, we may take into consideration the passive pilots used for the channel estimation of the cascaded CSI. To be specific, the reflecting elements of the IRS passively reflect the pilot sequences transmitted from the IoT devices to the AP/ PS to the IoT devices such that the CSI related to the IRS can be obtained \cite{LLiu_TWC2020}. In addition, while imperfect CSI will degrade the throughput performance due to the channel estimation errors, these channel uncertainties have been tackled in the IRS assisted wireless networks by taking into consideration robust resource allocations for active transmit and passive reflecting beamformers \cite{CPAN_Robust1_2020,CPAN_Robust2_2020,CPAN_Robust3_2020}. In particular, these robust schemes considered the imperfect cascaded BS-IRS-user channels, which typically follow  the bounded and statistical CSI uncertainties, and the worst-case and outage probability robust beamforming designs are considered.
	We denote $ \bm{\Theta}_{k} = \textrm{diag}\left[\beta_{k,1}\exp(j\alpha_{k,1}),...,\beta_{k,N}\exp(j\alpha_{k,N})\right],~| \exp(j\alpha_{k,n}) | = 1,~\forall k \in [0,K], n \in [1,N_{R}] $ as the diagonal matrices associated with the phase shifts of the IRS elements to reflect the $ k $-th IoT device, where $ \alpha_{k,n} \in [0, 2 \pi) $ and  $ \beta_{k,n} \in [0, 1] $ are the phase shift and amplitude of the associated reflection coefficient. When $ k = 0 $, $ \bm{\Theta}_{0} $ is known as the energy reflection phase shift matrix, whereas $ \bm{\Theta}_{k} $ is denoted as the information reflection phase matrix when $ k \in [1,K] $. Also, $ \beta_{k,n}, k \in [0,K], n \in [1,N_{R}] $ is typically set to be 1 to maximize the reflected signal at the $ n $-th element of the $ k $-th phase shift matrix. 
\subsection{Transmission Protocol}
In the IRS aided WPSN, we adopt a generic \emph{harvest-then-transmit} protocol, and the whole operation time period is set to $ T $. The PS first provides wireless energy to the IoT devices during the downlink WET duration $ \tau_{0} \in [0,1] $. At the same time, the IRS collects the energy signal and reflect it via IRS planar array. Then, these IoT devices utilize the harvested energy to independently transmit their own information to the AP with the help of the IRS by time division multiple access (TDMA). Thus, the uplink WIT time duration of the $ k $-th IoT device is denoted by $ \tau_{k} \in [0,1] $.
The time allocation satisfies $ \sum_{k=0}^{K} \tau_{k} = T $. The channel coefficients between the PS and the $ k $-th IoT device, the PS and the IRS, the IRS and the $ k $-th IoT device, the $ k $-th IoT device and the AP, the $ k $-th IoT device and the IRS, as well as the IRS and the AP are denoted by $ \mathbf{g}_{d,k} \in \mathbb{C}^{N_{T} \times 1} $, $ \mathbf{G}_{0} \in \mathbb{C}^{N_{R} \times N_{T}} $, $ \mathbf{g}_{r,k} \in \mathbb{C}^{N_{R} \times 1} $, $ h_{d,k} \in \mathbb{C}^{1 \times 1} $,  $ \mathbf{h}_{k} \in \mathbb{C}^{1 \times N_{R}} $, and $ \mathbf{h}_{r} \in \mathbb{C}^{N_{R} \times 1} $, respectively.\footnote{$ \mathbf{h}_{k} = \mathbf{g}_{r,k}^{T} $.} 
Thus, the RF harvested energy at the $ k $-th IoT device is given as\footnote{In this paper, we assume that all IoT devices employ the linear energy harvesting. This is due to the fact that this assumption practically holds when the harvested energy at the IoT devices is relatively lower than its battery capacity.}
\begin{align}
E_{k} = \eta \tau_{0} \left| \left(\mathbf{g}_{r,k}^{H} \bm{\Theta}_{0} \mathbf{G}_{0} + \mathbf{g}_{d,k}^{H}\right) \mathbf{w} \right|^{2},~\forall k \in [1,K], 
\end{align}
where $ \eta \in [0,1] $ is the energy conversion efficiency; $ \mathbf{w} $ denotes the energy beamforming of the PS satisfying $ \| \mathbf{w} \|^{2} \leq P_{0} $, $ P_{0} $ is the maximum transmit power available at the PS.
 It is assumed that all harvested energy at each IoT device is used for the information transmission, which achieves the maximum achievable throughput.\footnote{Practically, each IoT device also requires constant energy consumption to support its circuit operation. For convenience and without loss of generality, this circuit energy consumption can be assumed to be zero in our work.} Hence, the achievable throughput of the $ k $-th IoT device is given by 
\begin{align} 
&R_{k}(\bm{\tau}, \bm{\Theta})  \nonumber\\ & \!=\! \tau_{k} \log\! \left(\!\! 1 \!+\! \frac{\eta \tau_{0} \left| \left(\! \mathbf{g}_{r,k}^{H} \bm{\Theta}_{0} \mathbf{G}_{0} \!+\! \mathbf{g}_{d,k}^{H} \!\right) \mathbf{w} \right|^{2} \! | \mathbf{h}_{k} \bm{\Theta}_{k} \mathbf{h}_{r} \!+\! h_{d,k} |^{2}  }{\tau_{k} \sigma^{2}} \!\right)\!\!, \nonumber
\end{align}
where $ \bm{\tau} = [\tau_{0}, \tau_{1}, ..., \tau_{K}] $, $ \bm{\Theta} = [\bm{\Theta}_{0}, \bm{\Theta}_{1}, ..., \bm{\Theta}_{K}] $, and $ \sigma^{2} $ denotes the noise power at the AP.
\subsection{Problem Formulation}
To evaluate the performance of the IRS assisted WPSN, we are interested in maximizing the sum throughput which jointly designs the phase shifts of the WET and WIT phases as well as the transmission time allocation. Hence, the sum throughput maximization problem is formulated as
\begin{subequations}\label{eq:Original_problem_STM}
\begin{align}
\max_{\mathbf{w}, \bm{\tau},\bm{\Theta}} &~ \sum_{k=1}^{K}~ R_{k}(\bm{\tau}, \bm{\Theta}) 
\nonumber
\\
s.t.&~ | \exp\left( j \alpha_{k,n} \right) | \!=\! 1,~n \! \in \! [1,N_{R}], \label{eq:Original_problem_STM_phase} 
\\
&~ \sum_{k=0}^{K} \tau_{k} \! \leq \! T,~ \tau_{k} \! \geq \! 0,~\forall k \! \in \! [0,K],,  \label{eq:Original_problem_STM_time}  \\
&~ \| \mathbf{w} \|^{2} \leq P_{0}. \label{eq:Original_problem_STM_energy}
\end{align}
\end{subequations}
In problem \eqref{eq:Original_problem_STM}, \eqref{eq:Original_problem_STM_phase} is the constraint of phase shifts, \eqref{eq:Original_problem_STM_time} denotes the the transmission time constraint, and \eqref{eq:Original_problem_STM_energy} is the transmit power constraint of the PS. Problem \eqref{eq:Original_problem_STM} cannot be solved directly due to its non-convexity with respect to the energy beamforming $ \mathbf{w} $, the time allocation vector $ \bm{\tau} $ and the phase shift matrices $ \bm{\Theta} $. In the following, we first derive the phase shifts of the WIT phase in closed-form. Then, the SDP based AO and low complexity based AO schemes are applied to solve problem \eqref{eq:Original_problem_STM} which alternately designs the energy beamforming, the time allocation, and the phase shifts of the WET phase.
\section{Sum Throughput Maximization}\label{section:STM_IRS_WPSN}
In this section, we aim to solve the sum throughput maximization problem \eqref{eq:Original_problem_STM}. Specifically, we first optimize the phase shift matrix $ \bm{\Theta}_{k},~\forall k \in [1,K] $ by maximizing its objective function. To proceed, the following \emph{theorem} is presented to derive the optimal closed-form phase shifts of the WIT phase. 
	\begin{theorem}\label{lemma:Theta_k_problem}
		The optimal phase shifts of the WIT phases, i.e., $ \bm{\Theta}_{k} $ for $ \forall k \in [1,K]  $, are given as
		 \begin{align}
		 	\bm{\Theta}_{k}^{*} = \textrm{diag} \left( \theta_{k,1}^{*}, ..., \theta_{k,N_{R}}^{*} \right),~\forall k \in [1,K],~\forall n \in [1,N_{R}],
		 \end{align}
	 where $ \theta_{k,n}^{*} = \exp \left( j \alpha_{k,n}^{*} \right) $, $ \alpha_{k,n}^{*} = \arg \left( h_{d,k} \right) - \arg \left( \mathbf{b}_{k}[n] \right) $, and $ \mathbf{b}_{k} = \textrm{diag} \left( \mathbf{h}_{k} \right) \mathbf{h}_{r} $.
	\end{theorem}
\begin{IEEEproof}
See Appendix \ref{appendix:Lemma_Theta_k_problem}.
	\end{IEEEproof}  
For the optimal phase shift matrix $ \bm{\Theta}_{k}^{*} $, we have the following \emph{proposition}
\begin{proposition}\label{proposition:Theta_k}
The optimal phase shift matrix $ \bm{\Theta}_{k}^{*} $ aligns the cascaded link between the $ k $-th IoT device and the AP via the IRS with the direct link between them, which can be expressed as 
\begin{align}\label{eq:Relation_both_links}
\mathbf{h}_{k} \bm{\Theta}_{k}^{*} \mathbf{h}_{r} = \xi_{k} h_{d,k}, ~\forall k \in [1,K],
\end{align}
where $ \xi_{k} $ is positive scalar denoting the signal strength metric from the cascaded link between the $ k $-th IoT device and the AP via the IRS.
\end{proposition}
\begin{IEEEproof}
See Appendix \ref{appendix:Theta_k}.
	\end{IEEEproof}
By exploiting \emph{Proposition} \ref{proposition:Theta_k}, we have the following \emph{remark}
\begin{remark}
By exploiting \emph{Proposition} \ref{proposition:Theta_k}, for given $ \mathbf{w} $, $ \bm{\tau} $ and $ \bm{\Theta}_{0} $, the IRS can strengthen the received signal power of the $ k $-th IoT device at the AP at most $ (1 + \xi_{k})^{2} $ times in comparison to that without IRS. In addition, $ \xi_{k} $ for $ \forall k \in [1,K] $ is proportional to the number of the IRS reflecting elements. Hence, a significant improvement can be introduced in terms of the sum throughput performance with a larger number of reflecting elements. Moreover, the IRS deployment as well as the pathloss model of the cascaded link between the $ k $-th IoT device and the AP via the IRS also play an important role in affecting the received signal strength \cite{Bjornson_WCL_2020}. 
\end{remark}

To proceed, we define $ t_{k} = | \mathbf{h}_{k} \bm{\Theta}_{k}^{*} \mathbf{h}_{r} \!+\! h_{d,k} |^{2} $ and substitute it into problem \eqref{eq:Original_problem_STM} to get 
\begin{subequations}\label{eq:Problem_reformulation1}
\begin{align}
\max_{\mathbf{w},\bm{\tau},\bm{\Theta}_{0}}&~ \sum_{k=1}^{K} \tau_{k} \log \left( 1 + \frac{\eta \tau_{0} t_{k} \left| \left( \mathbf{g}_{r,k}^{H} \bm{\Theta}_{0} \mathbf{G}_{0} \!+\! \mathbf{g}_{d,k}^{H} \right) \mathbf{w} \right|^{2} }{\tau_{k} \sigma^{2} } \right)  \label{eq:Problem_reformulation1_obj}\\
s.t.&~ | \exp\left( j \alpha_{0,n} \right) | \!=\! 1,~ \forall n \in [1,N_{R}], \label{eq:Problem_reformulation1_phase}\\
&~\eqref{eq:Original_problem_STM_time},~\eqref{eq:Original_problem_STM_energy}.
\end{align}
\end{subequations}
Problem \eqref{eq:Problem_reformulation1} is still intractable due to the non-convex nature of its objective function \eqref{eq:Problem_reformulation1_obj} and the modulus constraint \eqref{eq:Problem_reformulation1_phase}. In order to circumvent this non-convex issue, we propose the AO algorithm to solve \eqref{eq:Problem_reformulation1}. Specifically, we first investigate the SDP relaxation scheme to reformulate \eqref{eq:Problem_reformulation1} into a convex optimization problem to alternately design the energy beamforming, the time allocation and the phase shifts of the WET phase. Consequently, we propose a low complexity scheme to derive the optimal closed-form energy beamforming, time allocation and the phase shifts of the WET phase in closed-form which efficiently reduces the computational complexity incurred by the SDP relaxation scheme.
\subsection{SDP Relaxation Scheme}\label{section:Global}
In this section, we consider an SDP relaxation scheme to alternately design the energy beamformer $ \mathbf{w} $, the time allocation $ \bm{\tau} $, and the phase shift matrix $ \bm{\Theta}_{0} $ by applying the AO algorithm \cite{QWu_TWC_2019,QWu_WCL_2020}. To be specific, we first fix $ \bm{\Theta}_{0} $ to optimize $ \mathbf{w} $ and $ \bm{\tau} $, then the $ \bm{\Theta}_{0} $ can be optimally designed for given $ \mathbf{w} $ and $ \bm{\tau} $. To proceed, we define $ \mathbf{V}_{0} = \tau_{0} \mathbf{W} $ and $ \mathbf{W} = \mathbf{w}\mathbf{w}^{H} $ and problem \eqref{eq:Problem_reformulation1} is relaxed as 
\begin{subequations}\label{eq:Problem_reformulation2}
	\begin{align}
	\max_{\bm{\tau},\mathbf{V}_{0} \succeq \mathbf{0}}&~ \sum_{k=1}^{K} \tau_{k} \log \left( 1 + \frac{ \tilde{t}_{k} \textrm{Tr}\left( \mathbf{\tilde{g}}_{k} \mathbf{\tilde{g}}_{k}^{H} \mathbf{V}_{0} \right)  }{\tau_{k} } \right)  \label{eq:Problem_reformulation2_obj}\\
	s.t. &~ \textrm{Tr}(\mathbf{V}_{0}) \leq \tau_{0} P_{0}, \label{eq:Problem_reformulation2_power}\\
	&~ \eqref{eq:Original_problem_STM_time},~\textrm{rank}(\mathbf{V}_{0}) = 1,  
	\end{align}
\end{subequations}
where $ \tilde{t}_{k} = \frac{ \eta t_{k} }{\sigma^{2}} $, $ \mathbf{\tilde{g}}_{k}^{H} =  \mathbf{g}_{r,k}^{H} \bm{\Theta}_{0} \mathbf{G}_{0} + \mathbf{g}_{d,k}^{H} $. The following \emph{lemma} is required to characterize the joint convexity of problem \eqref{eq:Problem_reformulation2}. 
\begin{lemma}\label{lemma:Convexity}
By dropping the non-convex rank-one constraint $ \textrm{rank}(\mathbf{V}_{0}) = 1 $, \eqref{eq:Problem_reformulation2} is jointly convex with respect to $ \bm{\tau} $ and $ \mathbf{V}_{0} $.
\end{lemma}
\begin{IEEEproof}
	See Appendix \ref{Appedix:Convexity}.
	\end{IEEEproof}

Note that the relaxed problem \eqref{eq:Problem_reformulation2} guarantees the rank-one nature of $ \mathbf{V}_{0} $, thus, the following \emph{lemma} is introduced to characterize the rank-profile of $ \mathbf{V}_{0} $.
\begin{lemma}\label{lemma:Rank_one_lemma}
	The optimal solution to problem \eqref{eq:Problem_reformulation2} is a rank-one matrix, i.e., $ \textrm{rank}(\mathbf{V}_{0}^{*}) = 1 $.
\end{lemma}
\begin{IEEEproof}
	See Appendix \ref{appendix:Rank_one}.
	\end{IEEEproof}
Since \eqref{eq:Problem_reformulation2} is a convex optimization problem without the non-convex rank-one constraint, we can employ interior-point methods \cite{boyd_B04}, to efficiently solve it to find the optimal solution $ (\mathbf{V}_{0}^{*}, \bm{\tau}^{*}) $. We further obtain $ \mathbf{W}^{*} = \frac{ \mathbf{V}_{0}^{*} }{ \tau_{0}^{*} } $ and $ \textrm{rank}(\mathbf{W}^{*}) = 1 $ by applying \emph{Lemma} \ref{lemma:Rank_one_lemma}. Thus, the optimal energy beamformer $ \mathbf{w}^{*} $ can be recovered from $ \mathbf{W}^{*} $ using eigen-decomposition.

Then, we design the phase shift matrix $ \bm{\Theta}_{0} $ for given $ \mathbf{w} $ and $ \bm{\tau} $. 
To proceed, problem \eqref{eq:Problem_reformulation1} is relaxed as 
\begin{subequations}\label{eq:Problem_reformulation3}
	\begin{align}
\max_{ \bm{\Delta}_{0} \succeq \mathbf{0} }&~ \sum_{k=1}^{K} \tau_{k} \log \left( 1 + \frac{ \tau_{0} \tilde{t}_{k} \textrm{Tr}\left( \bm{\Delta}_{0} \mathbf{F}_{k} \right)  }{\tau_{k} } \right)  \label{eq:Problem_reformulation3_obj}\\
s.t. &~ \bm{\Delta}_{0}(n,n) = 1,~\textrm{rank}(\bm{\Delta}_{0}) = 1,
\end{align}
\end{subequations} 
where $ \bm{\Delta}_{0} = \bm{\tilde{\theta}}_{0} \bm{\tilde{\theta}}_{0}^{H} $, $ \mathbf{F}_{k} = \mathbf{\tilde{D}}_{k} \mathbf{w} \mathbf{w}^{H} \mathbf{\tilde{D}}_{k}^{H} $, $ \bm{\tilde{\theta}}_{0}^{H} = \left[\!\!\begin{array}{cc}
	\bm{\theta}_{0}^{H} & 1
\end{array}\!\!\right] $, $ \bm{\theta}_{0} = [\exp(j \alpha_{0,1}), ..., \exp(j \alpha_{0,N_{R}})]^{H} $, and $ \mathbf{\tilde{D}}_{k} = \left[\!\! \begin{array}{cc}
	\textrm{diag}\left( \mathbf{g}_{r,k}^{H} \right)\mathbf{G}_{0} \\ \mathbf{g}_{d,k}^{H}
\end{array}\!\!\right] $. 
By dropping the non-convex rank-one constraint $ \textrm{rank}(\bm{\Delta}_{0}) = 1 $, \eqref{eq:Problem_reformulation3} is a convex optimization problem, which can be efficiently solved via interior-point methods \cite{boyd_B04}. 
Let us denote the solution to problem \eqref{eq:Problem_reformulation3} as $ \bm{\Delta}_{0}^{*} $. Due to the rank-one relaxation, the relaxed problem \eqref{eq:Problem_reformulation3} may have a higher rank solution, i.e., $ \textrm{rank}(\bm{\Delta}_{0}^{*}) \geq 1 $. In order to recover $ \tilde{\bm{\theta}}_{0} $ from $ \bm{\Delta}_{0}^{*} $, we apply a Gaussian randomization technique to construct an approximate rank-one solution \cite{QQWu_IRS_GLOBECOM_2018}. Specifically, the eigenvalue decomposition of $ \mathbf{V}_{0}^{*} $ is written as $ \bm{\Delta}_{0}^{*} = \bm{\Upsilon} \bm{\Lambda} \bm{\Upsilon}^{H} $, where $ \bm{\Upsilon} \in \mathbb{C}^{(N_{R}+1) \times (N_{R}+1)} $ is an
unitary matrix, and $ \bm{\Lambda} \in \mathbb{C}^{(N_{R}+1) \times (N_{R}+1)} $ is a diagonal matrix with eigenvalues arranged in decreasing order. To proceed, we construct a feasible solution as $ \tilde{\bm{\theta}}_{0}^{H} = \bm{\Upsilon} \bm{\Lambda}^{\frac{1}{2}} \bm{\kappa} $, where  $ \bm{\kappa} \in \mathbb{C}^{(N_{R}+1) \times 1} $ is randomly generated to follow complex circularly symmetric uncorrelated Gaussian variables with zero-mean and  covariance matrix $ \mathbf{I}_{N_{R}+1} $. According to \cite{QWu_TWC_2019}, the optimal phase vector is given by $ \bm{\theta}_{0}^{*} = [\exp( j \alpha_{0,1}^{*} ), ..., \exp( j \alpha_{0,N_{R}}^{*} ) ] $, where $ \alpha_{0,n}^{*} = \arg \left( \tilde{\bm{\theta}}_{0}(n) \right) $ for $ \forall n \in[1,N_{R}] $. Thus, the optimal $ \bm{\Theta}_{0}^{*} $ can be further obtained from $ \bm{\theta}_{0}^{*} $.
In the proposed AO algorithm, we replace solving problem \eqref{eq:Problem_reformulation1} with alternately solving problems \eqref{eq:Problem_reformulation2} and \eqref{eq:Problem_reformulation3} in an iterative fashion, in which the solution obtained in each iteration is used as the initial point of the next iteration. We summarize the detailed procedure of the proposed AO algorithm in \emph{Algorithm} \ref{algorithm:AO1}. 

\vspace{0.3em}
\hrule
\vspace{0.5em}
\begin{algorithm}\label{algorithm:AO1}
	\vspace{0.5em}
	The SDP based AO algorithm to solve problem \eqref{eq:Problem_reformulation1}.	
	\vspace{0.5em}
	\hrule
		\vspace{0.5em}
	\begin{enumerate}
		\item \textbf{Initialization}: IRS phase shift matrix $ \bm{\Theta}_{0}^{(0)} $ with iteration number $ m $.
		\item \textbf{Repeat}: Iteration number $ m $
		\begin{enumerate}
			\item \textbf{Fix} $ \bm{\Theta}_{0} = \bm{ \Theta }_{0}^{(m)} $ to obtain the optimal solution of the energy beamforming $ \mathbf{w}^{(m+1)} $ and the time allocation $ \bm{\tau}^{(m+1)} $ by solving problem \eqref{eq:Problem_reformulation2}.
			\item \textbf{Fix} $ \mathbf{w} = \mathbf{w}^{(m+1)} $ and $ \bm{\tau} = \bm{\tau}^{(m+1)} $ to obtain the optimal solution of the phase shift matrix $ \bm{\theta}_{0}^{(m+1)} $ by solving problem \eqref{eq:Problem_reformulation3}.
			\item \textbf{Set} $ \bm{ \Theta}_{0}^{(m+1)} = \textrm{diag}\left( \bm{\theta}_{0}^{(m+1)} \right) $.
			\item \textbf{Update}: $ m = m + 1 $
		\end{enumerate}
	\item \textbf{Until convergence.}
	\end{enumerate}
\end{algorithm}
\vspace{0.5em}
\hrule
\vspace{0.5em}

Now, the following \emph{proposition} is required to characterize the convergence of \emph{Algorithm} \ref{algorithm:AO1}
\begin{proposition}\label{proposition:Convergence1}
	The objective value of problem \eqref{eq:Problem_reformulation2} produces a non-descending trend over iterations by employing \emph{Algorithm} \ref{algorithm:AO1}, and its convergence is guaranteed.
\end{proposition}
\begin{IEEEproof}
	See Appendix \ref{appendix:Convergence1}.
	\end{IEEEproof}

Note that the maximum values of problems\eqref{eq:Problem_reformulation2} and \eqref{eq:Problem_reformulation3} serve as upper bounds for the reformulated counterparts of problem \eqref{eq:Problem_reformulation1} with respect to $ \mathbf{w} $ for given $ \bm{\tilde{\theta}}_{0} $ and with respect to $ \bm{\tilde{\theta}}_{0} $ for given $ \mathbf{w} $, respectively \cite{XYu_JSAC_2020}.
	We solve problems \eqref{eq:Problem_reformulation2} and \eqref{eq:Problem_reformulation3} in Steps (2-a) and (2-b) of \emph{Algorithm} \ref{algorithm:AO1} in an iterative fashion such that these upper bounds can be monotonically tightened, and the objective values of each iteration $  (\mathbf{w}^{(m)}, \bm{\tau}^{(m)}, \bm{\Theta}_{0}^{(m)})$ yield a non-decreasing sequence that converges to a stationary value in polynomial time. Any limit point of the sequence $  (\mathbf{w}^{(m)}, \bm{\tau}^{(m)}, \bm{\Theta}_{0}^{(m)})$ is a stationary point of problem \eqref{eq:Problem_reformulation1} \cite{Palomar_TSP2017,XYu_JSAC_2020}. Also, provided that problem \eqref{eq:Problem_reformulation1} is always feasible, \emph{Algorithm} \ref{algorithm:AO1} yields a feasible solution \cite{HSHEN_CL_2019}.

\subsection{Low Complexity Scheme}\label{section:Low_complexity}
In Section \ref{section:Global}, we proposed the SDP based AO scheme to solve problem \eqref{eq:Problem_reformulation1}, which alternately designs the energy beamforming, the time allocation, and the phase shifts of the WET phase. However, the AO base SDP scheme may incur a very high computational complexity due to the SDP relaxation with Gaussian randomization.
Thus, it is imperative to develop an independent scheme which can be deployed on the IRS and the IoT devices. In this subsection, we propose the low complexity based AO scheme which derives an optimal closed-form solution to alternately solve problem \eqref{eq:Problem_reformulation1} to significantly reduce the computation complexity compared with the SDP based AO scheme. 
\subsubsection{Optimization of $ \mathbf{w} $ and  {\boldmath $\tau$} for given {\boldmath $ \Theta_{0} $}}\label{section:w_tau}
First, let $ \mathbf{W} = \mathbf{w} \mathbf{w}^{H} $, we rewrite problem \eqref{eq:Problem_reformulation1} for given $ \bm{\Theta}_{0} $ as \eqref{eq:Problem_tau_W} \footnote{Here, problem \eqref{eq:Problem_reformulation1} is rewritten as \eqref{eq:Problem_tau_W} instead of \eqref{eq:Problem_reformulation2}. This is due to the fact that it is more convenient to derive closed-form solutions of the transmission time allocation and the energy beamforming. Note that both problems \eqref{eq:Problem_reformulation2} and \eqref{eq:Problem_tau_W} have the same solution. }
\begin{subequations}\label{eq:Problem_tau_W}
\begin{align}
\max_{\mathbf{W} \succeq \mathbf{0},\bm{\tau}}&~ \sum_{k=1}^{K} \tau_{k} \log \left( 1 + \frac{\tau_{0} \tilde{t}_{k}  \textrm{Tr}(\mathbf{\tilde{g}}_{k} \mathbf{\tilde{g}}_{k}^{H} \mathbf{W} )    }{\tau_{k} } \right)  \label{eq:Problem_tau_W_obj}  \\
s.t.&~
 \textrm{Tr} (\mathbf{W}) \leq P_{0},~\eqref{eq:Original_problem_STM_time}.
\end{align}
\end{subequations}

Note that $ \textrm{rank}(\mathbf{W}) = 1 $ is satisfied in problem \eqref{eq:Problem_tau_W} due to the SDP relaxation (i.e., $ \mathbf{W} = \mathbf{w} \mathbf{w}^{H} $). Problem \eqref{eq:Problem_tau_W} is not jointly convex with respect to $ \mathbf{W} $ and $ \bm{\tau} $ due to the non-convex property of \eqref{eq:Problem_tau_W_obj} which cannot be directly solved. 
In order to circumvent this issue, we consider the following Lagrange dual function for given $ \mathbf{W} $ to derive a closed-form solution for $ \bm{\tau} $.
\begin{align}\label{eq:Lagrange_function_tau}
\mathcal{L} (\bm{\tau},\mu) &~= \sum_{k=1}^{K} \tau_{k} \log \left( 1 + \frac{ \tau_{0}  \tilde{t}_{k}  \textrm{Tr}(\mathbf{\tilde{g}}_{k} \mathbf{\tilde{g}}_{k}^{H} \mathbf{W} )    }{\tau_{k} } \right)  \nonumber\\ &~~~~ - \mu \left( \tau_{0} + \sum_{k=1}^{K} \tau_{k} - T \right), 
\end{align}
where $ \mu $ is the non-negative dual multiplier associated with the constraint \eqref{eq:Original_problem_STM_time}. Also, its dual problem is given by
\begin{align}
\min_{ \bm{\tau} \in \mathcal{S} } ~ \mathcal{L} (\bm{\tau},\mu),
\end{align}
where $ \mathcal{S} $ is the feasible set of any $ \tau_{k} $ for $ \forall k \in [0,K] $, which has been shown in the constraint \eqref{eq:Original_problem_STM_time}. Note that \eqref{eq:Problem_tau_W} can be reformulated into a convex optimization problem and guarantees Slater’s condition, since $ \bm{\tau} \in \mathcal{S} $, $ \bm{\tau} \succeq \mathbf{0} $, and $ \sum_{k=1}^{K} \tau_{k} < T $. Hence, strong duality holds such that the optimal solution to \eqref{eq:Problem_tau_W} satisfies the Karush-Kuhn-Tucker (KKT) conditions \cite{boyd_B04}, which is given by
\begin{subequations}
	\begin{align}
	&~ \mu \left( \tau_{0} + \sum_{k=1}^{K} \tau_{k} - T \right) = 0, \label{eq:KKT1} \\
	&~ \frac{ \partial \mathcal{L} }{ \partial \tau_{k} } = 0,~\forall k \in [1,K]. \label{eq:KKT2}
	\end{align}
\end{subequations}
From \eqref{eq:KKT1}, we have the optimal dual variable satisfies $ \mu^{*} > 0 $, since the optimal solution to problem \eqref{eq:Problem_tau_W} (i.e., $ \bm{\tau}^{*} $) guarantees the equality $ \sum_{k = 0}^{K} \tau_{k} = T $.
According to \eqref{eq:KKT2}, we take into consideration the first-order derivative of \eqref{eq:Lagrange_function_tau} in terms of $ \tau_{k} $ for $ \forall k \in [1,K] $ and set it to zero, which is given by
\begin{align}\label{eq:First_derivative_tau_k}
\!\!\!\!\! \log \left(\!  1 \!+\! \frac{ \tau_{0}  \tilde{t}_{k}  \textrm{Tr}(\mathbf{\tilde{g}}_{k} \mathbf{\tilde{g}}_{k}^{H} \mathbf{W} )    }{\tau_{k} }  \!\right) \!-\! \frac{ \tau_{0}  \tilde{t}_{k}  \textrm{Tr}(\mathbf{\tilde{g}}_{k} \mathbf{\tilde{g}}_{k}^{H} \mathbf{W} )  }{ \tau_{k} +  \tau_{0}  \tilde{t}_{k}  \textrm{Tr}(\mathbf{\tilde{g}}_{k} \mathbf{\tilde{g}}_{k}^{H} \mathbf{W} )  }  \!=\!  \mu.\!\!\!\!
\end{align}
The left hand side (LHS) of \eqref{eq:First_derivative_tau_k} can be written in the form of $ f(x) = \log (1 + x) - \frac{x}{1 + x} $, which is a monotonically increasing function with respect to $ x $. To guarantee the $ K $ equations in \eqref{eq:First_derivative_tau_k}, we further have 
\begin{align}
\frac{ \tau_{0}  \tilde{t}_{1} \textrm{Tr} ( \mathbf{\tilde{g}}_{1} \mathbf{\tilde{g}}_{1}^{H} \mathbf{W} ) }{ \tau_{1} } 
\!=\! ... \!=\! \frac{ \tau_{0}  \tilde{t}_{K} \textrm{Tr} ( \mathbf{\tilde{g}}_{K} \mathbf{\tilde{g}}_{K}^{H} \mathbf{W} ) }{ \tau_{K} }.
\end{align}
Let $ \frac{1}{\lambda} = \frac{ \tau_{0}  \tilde{t}_{k} \textrm{Tr} ( \mathbf{\tilde{g}}_{k} \mathbf{\tilde{g}}_{k}^{H} \mathbf{W} ) }{ \tau_{k} } $, we have
\begin{align}\label{eq:tau_k1}
\tau_{k} = \lambda \tau_{0}  \tilde{t}_{k} \textrm{Tr} ( \mathbf{\tilde{g}}_{k} \mathbf{\tilde{g}}_{k}^{H} \mathbf{W} ),
\end{align}
Substitute \eqref{eq:tau_k1} into the constraint \eqref{eq:Original_problem_STM_time}, the following derivations are required to obtain the following expression for $ \lambda $ 
\begin{align}\label{eq:Lambda_expression}
\lambda = \frac{ T - \tau_{0} }{ \sum_{k =1 }^{K} \tau_{0}   \tilde{t}_{k} \textrm{Tr} ( \mathbf{\tilde{g}}_{k} \mathbf{\tilde{g}}_{k}^{H} \mathbf{W} ) }.
\end{align}
Substitute \eqref{eq:Lambda_expression} into \eqref{eq:tau_k1}, the optimal solution of $ \tau_{k} $ for $ \forall k \in [1,K] $ is given by 
\begin{align}\label{eq:Tauk}
\tau_{k}^{*} = \frac{ ( T - \tau_{0} ) \tau_{0} \tilde{t}_{k} \textrm{Tr} ( \mathbf{\tilde{g}}_{k} \mathbf{\tilde{g}}_{k}^{H} \mathbf{W} ) }{ \sum_{k =1 }^{K} \tau_{0} \tilde{t}_{k} \textrm{Tr} ( \mathbf{\tilde{g}}_{k} \mathbf{\tilde{g}}_{k}^{H} \mathbf{W} ) }.
\end{align}
We substitute \eqref{eq:Tauk} into problem \eqref{eq:Problem_tau_W} and let $ \mathbf{\tilde{G}} = \left[ \sqrt{\tilde{t}_{1}} \mathbf{\tilde{g}}_{1}, ..., \sqrt{\tilde{t}_{K}} \mathbf{\tilde{g}}_{K} \right] $, \eqref{eq:Problem_tau_W} is written as 
\begin{subequations}\label{eq:IRS_WPSN_problem_reformulation2}
\begin{align}
\max_{ \mathbf{W} \succeq \mathbf{0},\tau_{0}} &~ ( T - \tau_{0} )  \log \left( 1 + \frac{ \tau_{0}  \textrm{Tr} \left( \mathbf{\tilde{G}} \mathbf{\tilde{G}}^{H} \mathbf{W}  \right) }{ T - \tau_{0} } \right)  \label{eq:IRS_WPSN_problem_reformulation2_obj}\\
s.t.&~  \textrm{Tr}(\mathbf{W}) \leq  P_{0},~ 0 \leq \tau_{0} \leq T.
\end{align}
\end{subequations}
Problem \eqref{eq:IRS_WPSN_problem_reformulation2} is still not jointly convex in terms of $ \mathbf{W} $ and $ \tau_{0} $. In the following, we derive the optimal closed-form solution $ \mathbf{W} $ for given $ \tau_{0} $ using the following \emph{theorem}.
\begin{theorem}\label{theorem:Closedform_W}
	The optimal closed-form solution to problem \eqref{eq:IRS_WPSN_problem_reformulation2}, i.e., $ \mathbf{W} $, is given by $ \mathbf{W}^{*} = P_{0} \bm{\nu}_{\max} \left(\mathbf{\tilde{G}} \mathbf{\tilde{G}}^{H} \right) \bm{\nu}_{\max} \left(\mathbf{\tilde{G}} \mathbf{\tilde{G}}^{H} \right)^{H}
	$. 
\end{theorem}
\begin{IEEEproof}
	See Appendix \ref{appendix:Closedform_W}.
	\end{IEEEproof}
Due to the rank-one condition $ \textrm{rank}(\mathbf{W}) = 1 $, the optimal energy beamformer is easily derived as $ \mathbf{w}^{*} = \sqrt{P_{0}} \bm{\nu}_{\max} \left( \mathbf{\tilde{G}} \mathbf{\tilde{G}}^{H} \right) $. By applying \emph{Theorem} \ref{theorem:Closedform_W}, problem \eqref{eq:IRS_WPSN_problem_reformulation2} can be rewritten with respect to $ \tau_{0} $ as 
\begin{align}\label{eq:Subproblem_tau0}
\max_{\tau_{0}} &~ (T - \tau_{0}) \log \left( 1 + \frac{\tau_{0} P_{0} \rho_{\max} \left( \mathbf{\tilde{G}} \mathbf{\tilde{G}}^{H} \right) }{ T - \tau_{0} } \right) \nonumber\\ 
s.t.&~ 0 \leq \tau_{0} \leq T.
\end{align}
Problem \eqref{eq:Subproblem_tau0} is a single-value optimization problem, where the optimal time allocation $ \tau_{0} $ can be easily attained by using one-dimensional line search, e.g., golden search \cite{QSun_CL_WPCN_2014}. However, this numerical search will introduce a very high computational complexity. In order to circumvent this issue, we propose a low complexity method to derive the optimal closed-form $ \tau_{0} $ without such an exhaustive search. To proceed, the following \emph{theorem} is required
\begin{theorem}\label{theorem:Optimal_tau0}
	The optimal time allocation $ \tau_{0} $ can be derived in closed-form as
	\begin{align}\label{eq:Optimal_tau0}
	\tau_{0}^{*} \!=\! \frac{ \exp \left[ \mathcal{W} \left( \frac{ P_{0} \rho_{\max} \left(\! \mathbf{\tilde{G}} \mathbf{\tilde{G}}^{H} \!\right) \!-\! 1 }{ \exp(1) } \right) \!+\! 1 \right]  \!-\! 1 }{  P_{0} \rho_{\max} \left(\! \mathbf{\tilde{G}} \mathbf{\tilde{G}}^{H} \!\right) \!-\! 1 \!+\!  \exp \left[ \mathcal{W} \left( \frac{ P_{0} \rho_{\max} \left( \mathbf{\tilde{G}} \mathbf{\tilde{G}}^{H} \right) \!-\! 1 }{ \exp(1) }  \right) \!+\! 1 \right]  } T.
	\end{align}
\end{theorem}
\begin{IEEEproof}
	See Appendix \ref{appendix:Optimal_tau0}.
	\end{IEEEproof}

\subsubsection{Optimization of {\boldmath $ \Theta_{0} $} for given $ \mathbf{w} $ and  {\boldmath $\tau$}}
Here, we first rewrite problem \eqref{eq:Problem_reformulation1} with respect to $ \bm{\Theta}_{0} $ as
\begin{align}\label{eq:Problem_reformulation12}
\max_{ \bm{\Theta}_{0} } &~ ( 1 \!-\! \tau_{0} )  \log \left(\! 1 \!+\! \frac{ \tau_{0} \sum_{k =1 }^{K} \tilde{t}_{k} \left | \left( \mathbf{g}_{r,k}^{H} \bm{\Theta}_{0} \mathbf{G}_{0} \!+\! \mathbf{g}_{d,k}^{H}  \right) \mathbf{w} \right |^{2} }{ 1 \!-\! \tau_{0} } \right) \nonumber\\
s.t.&~
| \exp\left( j \alpha_{0,n} \right) | \!=\! 1,~ \forall n \in [1,N_{R}], 
\end{align}
Problem \eqref{eq:Problem_reformulation12} is not convex due to its non-convex unit-modulus constraint. Thus, we propose a novel method to derive the optimal closed-form phase shifts of the WET phase. It is easily verified that maximizing problem \eqref{eq:Problem_reformulation12} is equivalent to the maximization of the following problem 
\begin{subequations}\label{eq:Problem_reformulation14}
\begin{align}
\max_{ \bm{\theta}_{0} } &~ \sum_{k =1 }^{K} \tilde{t}_{k} \left|  \bm{\theta}_{0}^{H} \mathbf{b}_{k}  + d_{k}  \right|^{2} , \label{eq:Problem_reformulation14_obj}    \\
s.t.&~
| \bm{\theta}_{0}(n) | \!=\! 1,~ \forall n \in [1,N_{R}],
\end{align}
\end{subequations}
where $ \mathbf{b}_{k} = \textrm{diag}\left( \mathbf{g}_{r,k}^{H} \right)\mathbf{G}_{0} \mathbf{w} $ and $ d_{k} = \mathbf{g}_{d,k}^{H}\mathbf{w} $.
By applying a few mathematical manipulations to \eqref{eq:Problem_reformulation14_obj}, we have 
\begin{align}
\sum_{k =1 }^{K} \tilde{t}_{k} \left|  \bm{\theta}_{0}^{H} \mathbf{b}_{k}  + d_{k}  \right|^{2}  
&~=  \bm{\theta}_{0}^{H}  \bm{\Phi}_{1}  \bm{\theta}_{0} + 2 \Re \left\{ \bm{\theta}_{0}^{H} \bm{\gamma} \right\} +  e_{1},
\end{align}
where $ \bm{\Phi}_{1} = \sum_{k=1}^{K} \tilde{t}_{k} \mathbf{b}_{k}\mathbf{b}_{k}^{H} $, $ \bm{\gamma} = \sum_{k=1}^{K} \tilde{t}_{k} \textrm{conj}(d_{k}) \mathbf{b}_{k} $, and $ e_{1} = \sum_{k=1}^{K}  \tilde{t}_{k} d_{k} \textrm{conj}(d_{k}) $.
Accordingly, problem \eqref{eq:Problem_reformulation14} is equivalent to 
\begin{subequations}\label{eq:Problem_reformulation15}
\begin{align}
\min_{\bm{\theta}_{0}} &~ \bm{\theta}_{0}^{H}  \bm{\Phi}  \bm{\theta}_{0} - 2 \Re \left\{ \bm{\theta}_{0}^{H} \bm{\gamma} \right\} +  e, \label{eq:Problem_reformulation15_obj}  \\
s.t.&~ | \bm{\theta}_{0}(n) | = 1, ~\forall n \in [1, N_{R}], \label{eq:Problem_reformulation15_constraint}
\end{align}
\end{subequations}
where
$ \bm{\Phi} = - \bm{\Phi}_{1} $, 
and $ e = -e_{1} $.

Problem \eqref{eq:Problem_reformulation15} is still intractable due to its unit modulus equality constraint \eqref{eq:Problem_reformulation15_constraint}.
In order to tackle this issue, the Majorization-Minimization (MM) algorithm is adopted, and a sequence of tractable sub-problems are considered to iteratively solve problem \eqref{eq:Problem_reformulation15} by approximating its objective function and constraint set \cite{Palomar_TSP2017}.
 Now, consider the following problem: 
\begin{align}\label{eq:MM_problem}
\min_{\mathbf{x}} &~ f_{0}(\mathbf{x}),~
s.t.~ f_{i}(\mathbf{x}) \leq 0, ~i = 1, ..., L.
\end{align} 
We approximate both the objective function and feasible constraints set of problem \eqref{eq:MM_problem} at each iteration.\footnote{Here we assume that $ f_{i} $ is differential \cite{Palomar_TSP2017}.} Thus, the following convex sub-problem can be solved at the $ m $-th iteration. 
\begin{align}\label{eq:Subproblem_MM}
\min_{\mathbf{x}}&~ g_{0} (\mathbf{x}|\mathbf{x}^{(m)}) ,~
s.t.~ g_{i} (\mathbf{x}|\mathbf{x}^{(m)}) \leq 0, i = 1, ..., L,
\end{align} 
where $ g_{i} (\mathbf{*}|\mathbf{x}^{(m)}) $, $ \forall m = 0, ..., L $ denotes a convex function which guarantees the following conditions: 
\begin{align}\label{eq:MM_three_conditions}
g_{i}(\mathbf{x}^{(m)}|\mathbf{x}^{(m)})& = f_{i}(\mathbf{x}^{(m)}), \nonumber\\
g_{i}(\mathbf{x}|\mathbf{x}^{(m)})& \geq f_{i}(\mathbf{x}), \\
\nabla g_{i} (\mathbf{x}^{(m)}|\mathbf{x}^{(m)}) & = \nabla f_{i} (\mathbf{x}^{(m)}). \nonumber
\end{align}
The sequence $ \mathbf{x}^{(m)} $ results in a monotonically decreasing $ f_{0}(\mathbf{x}) $ which converges to a KKT point \cite{Palomar_TSP2017}. In other words, the sub-problem \eqref{eq:Subproblem_MM} is introduced from the upper bound of the objective function in \eqref{eq:MM_problem} via a convex surrogate function, and the feasible set in \eqref{eq:MM_problem} is approximated via linearization \cite{Palomar_TSP2017}.

\begin{proposition}\label{proposition:Surrogate_MM}\cite{Palomar_TSP_2016,Palomar_TSP2017}
	The objective function \eqref{eq:Problem_reformulation15_obj} is approximated in the following for any given $ \bm{\theta^{(m)}} $ at the $ m $-th iteration and for any feasible $ \bm{\theta} $.
	\begin{align}\label{eq:Surrogate_function_MM}
& ~ 	f(\bm{\theta}_{0}) \triangleq ~ \bm{\theta}_{0} \bm{\Phi} \bm{\theta}_{0}^{H} \!-\! 2 \Re \{ \bm{\theta}_{0} \bm{\gamma} \} \!+\! e \nonumber\\ 
	&	\leq \bm{\theta}_{0} \bm{\Upsilon} \bm{\theta}_{0}^{H} \!-\! 2\Re \left\{ \bm{\theta}_{0} \left[ (\bm{\Upsilon} - \bm{\Phi}) \tilde{\bm{\theta}}_{0}^{H} \!+\! \bm{\gamma} \right] \right\} 
	\!+\!  \tilde{\bm{\theta}}_{0}  (\bm{\Upsilon} \!-\! \bm{\Phi}) \tilde{\bm{\theta}}_{0}^{H} \!+\!  e\nonumber\\
	& \!=\! \rho_{ \max } ( \bm{\Phi} ) \| \bm{\theta}_{0} \|^{2} \!-\! 2 \mathcal{R} \left\{\!  \bm{\theta}_{0} \left[ \left(\! \rho_{ \max } ( \bm{\Phi} ) \mathbf{I}_{N_{R} \times N_{R}} \!-\! \bm{\Phi} \! \right) \tilde{\bm{\theta}}_{0}^{H} 
	\!+\! \bm{\gamma}  \right] \!\right\} \nonumber\\ &~~~~  + \tilde{e}  
	~\triangleq ~g(\bm{\theta}|\bm{\theta}^{(m)}),
	\end{align}
	where $ \tilde{e} =  \tilde{\bm{\theta}}_{0} \left[ \rho_{ \max } ( \bm{\Phi} ) \mathbf{I}_{N_{R} \times N_{R}} - \bm{\Phi} \right] \tilde{\bm{\theta}}_{0}^{H} + e $, and $ \bm{\Upsilon} = \rho_{\max}( \bm{\Phi} ) \mathbf{I}_{N_{R} \times N_{R}} $. 
	Also, $ \tilde{\bm{\theta}}_{0} $ denotes the approximated solution to $ \bm{\theta}_{0} $ which is achieved in the previous iteration via the alternating algorithm. 
\end{proposition}
\emph{Proposition} \ref{proposition:Surrogate_MM} constructs a surrogate function of \eqref{eq:Problem_reformulation15_obj}, and it is easily verified that $ g (\bm{\theta}|\bm{\theta}^{(m)}) $ in \eqref{eq:Surrogate_function_MM} guarantees the conditions in \eqref{eq:MM_three_conditions}. By exploiting \emph{Proposition} \ref{proposition:Surrogate_MM}, problem \eqref{eq:Problem_reformulation15} is reformulated as 
\begin{align}\label{eq:Problem_reformulation16}
\min_{\bm{\theta}_{0}} &~ \rho_{\max} \left( \bm{\Phi} \right) \|\bm{\theta}_{0}\|^{2} - 2 \Re \left\{ \bm{\theta}_{0}^{H} \tilde{\bm{\gamma}} \right\}, \nonumber\\
s.t.&~ | \bm{\theta}_{0}(n) | = 1,~\forall n \in [1,N_{R}],
\end{align}
where $ \tilde{\bm{\gamma}} =   \left( \rho_{ \max } ( \bm{\Phi} ) \mathbf{I}_{N_{R} \times N_{R}} - \bm{\Phi} \right) \tilde{\bm{\theta}}_{0}^{H} + \bm{\gamma}  $.
Also, it is easy to obtain $ \| \bm{\theta}_{0} \|^{2} = N_{R} $ due to $  | \bm{\theta}_{0}(n) | = 1,~\forall n \in [1,N_{R}] $.
The term $ \Re \left\{ \bm{\theta}_{0} \tilde{\bm{\gamma}} \right\} $ can be maximized when the phases of $ \bm{\theta}_{0}(n) $ and $ \tilde{\bm{\gamma}}(n) $ are identical. Hence, the optimal solution to \eqref{eq:Problem_reformulation16} is derived as 
\begin{align}\label{eq:Closed_form_theta0}
\bm{\theta}_{0}^{*} = \left[\!\!\begin{array}{ccc}
\exp \left(j \arg [\tilde{\bm{\gamma}}(1)] \right) \!\!& ,..., & \!\!\exp \left(j \arg [\tilde{\bm{\gamma}}(N)] \right)
\end{array}\!\!\right].
\end{align}
Although problem \eqref{eq:Problem_reformulation14} can be easily relaxed as a SDP, the optimal closed-form phase shift in \eqref{eq:Closed_form_theta0} provides a more efficient way for practical implementation and reduction of the computational complexity introduced by the SDP, especially for a large number of reflecting elements $ N_{R} $.

\vspace{0.5em}
\hrule
\vspace{0.5em}
\begin{algorithm}\label{algorithm:AO2}
	\vspace{0.5em}
	The low complexity based AO algorithm to solve problem \eqref{eq:Problem_reformulation1}.	
		\vspace{0.5em}
	\hrule
	\vspace{0.5em}
	\begin{enumerate}
		\item \textbf{Initialization}: IRS phase shift matrix $ \bm{\Theta}_{0}^{(0)} $ with iteration number $ m $.
		\item \textbf{Repeat}: Iteration number $ m $
		\begin{enumerate}
			\item \textbf{Fix} $ \bm{\Theta}_{0} = \bm{\Theta}_{0}^{(m)} $ to optimize energy beamforming $ \mathbf{w}^{(m+1)} $ and time allocation $ \bm{\tau}^{(m+1)} = \left[\tau_{0}^{(m+1)}, ... , \tau_{K}^{(m+1)}\right] $.
			\begin{enumerate}
				\item \textbf{Obtain} $ \mathbf{w}^{(m+1)} $ via \emph{Theorem} \ref{theorem:Closedform_W}.
				\item \textbf{Obtain} $ \tau_{0}^{(m+1)}$ via \emph{Theorem} \ref{theorem:Optimal_tau0}.
				\item \textbf{Obtain} $ \tau_{k}^{(m+1)} $ for $ \forall k \in [1,K] $ via \eqref{eq:Tauk}.
			\end{enumerate}
			\item \textbf{Fix} $ \mathbf{w} = \mathbf{w}^{(m+1)} $ and $ \bm{\tau} = \bm{\tau}^{(m+1)} $ to optimize phase shift vector $ \bm{\theta}_{0}^{(m+1)} $ which can be obtained via \eqref{eq:Closed_form_theta0}.
			\item \textbf{Set} $ \bm{\Theta}_{0}^{(m+1)} = \textrm{diag}\left(\bm{\theta}_{0}^{(m+1)}\right) $.
			\item \textbf{Update} $ m = m + 1 $.
		\end{enumerate}
		\item \textbf{Until convergence.}
	\end{enumerate}
\end{algorithm}
\vspace{0.5em}
\hrule
\vspace{0.5em}
The objective value of problem \eqref{eq:Problem_tau_W} produces a non-decreasing trend over iterations by applying \emph{Algorithm} \ref{algorithm:AO2}, and its convergence is guaranteed. The proof of this statement is similar to that of \emph{Proposition} \ref{proposition:Convergence1}, and is omitted here due to space limitation.
\section{Numerical Results}\label{section:simulation}
\begin{figure}[!htbp]
	\centering
	\includegraphics[scale = 0.6]{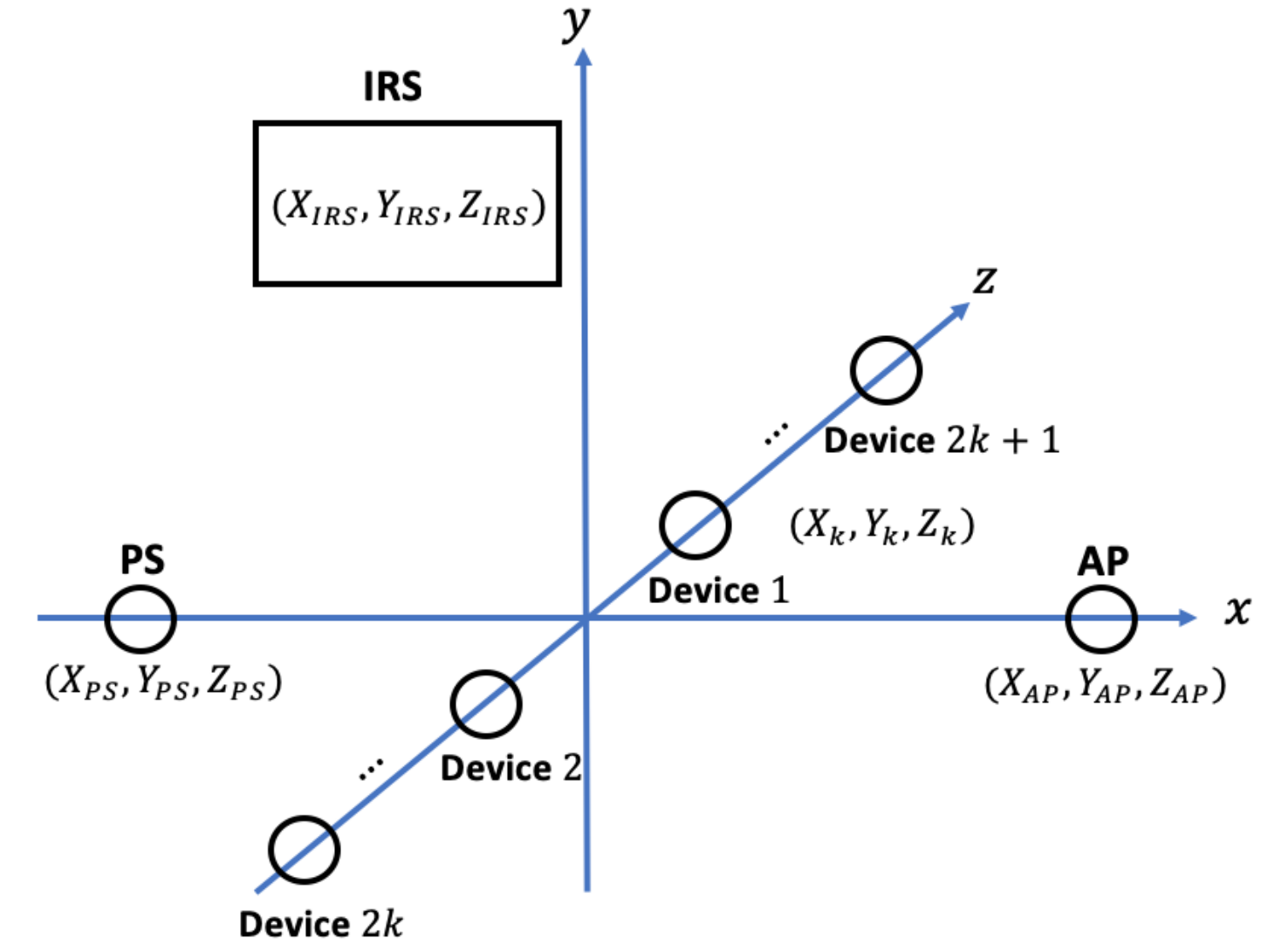}
	\caption{System deployment.}
	\label{fig:System_deployment}
\end{figure}
In this section, numerical results are presented to illustrate the theoretical results of the proposed algorithms as derived in Section \ref{section:STM_IRS_WPSN}. In our simulations, we consider the system deployment shown in Fig. \ref{fig:System_deployment}. Unless otherwise stated, the PS, the AP, and the IRS are located at $ (-10,0,0) $, $ (10,0,0) $ and $ (-2,6,0) $. All IoT sensor nodes are assigned at $ (0,0, \frac{n\times l}{2}) $ if $ n = 1,...,(2k+1) $, and $ \left(0,0,-\frac{(n-1)\times l }{2}\right) $, if $ n = 2,...,2k $, where $ l $ is the interval between two neighbouring IoT devices. The channel coefficient is composed of distance-dependent path loss model and small-scale fading. The path loss model is set to $ P_{L} = A d^{ -\varepsilon } $, where $ A = -30~\textrm{dB} $, $ \varepsilon $ is the path loss exponent, and $ d $ represents the distance between any two nodes, i.e., the PS and the IRS, the PS and the $ k $-th IoT devices, the IRS and the $ k $-th IoT device, the IRS and the AP, as well as the $ k $-th IoT device and the AP.
The channel coefficients between the PS and the IRS, the IRS and the $ k $-th IoT device, the $ k $-th IoT device and the IRS, as well as the IRS and the AP are modelled as $ \mathbf{G}_{0} = \sqrt{ \frac{K_{1}}{K_{1}+1} } \mathbf{G}_{0}^{\textrm{LOS}} + \sqrt{\frac{1}{K_{1}+1}} \mathbf{G}_{0}^{\textrm{NLOS}} $, $ \mathbf{g}_{r,k} = \sqrt{ \frac{K_{1}}{K_{1}+1} }  \mathbf{g}_{r,k}^{\textrm{LOS}} +  \sqrt{ \frac{ 1 }{ K_{1}+1} } \mathbf{g}_{r,k}^{\textrm{NLOS}} $,
$ \mathbf{h}_{k} = \mathbf{g}_{r,k}^{T} $, 
and $ \mathbf{h}_{r} = \sqrt{ \frac{K_{1}}{K_{1}+1} } \mathbf{h}_{r}^{\textrm{LOS}} + \sqrt{ \frac{ 1 }{ K_{1}+1} } \mathbf{h}_{r}^{\textrm{NLOS}} $,
where $ \mathbf{G}_{0}^{\textrm{LOS}} $, $ \mathbf{g}_{r,k}^{\textrm{LOS}} $, and $ \mathbf{h}_{r}^{\textrm{LOS}} $ 
denote the line-of-sight (LOS) deterministic components of the corresponding channel coefficients; $ \mathbf{G}_{0}^{\textrm{NLOS}} $, $ \mathbf{g}_{r,k}^{\textrm{NLOS}} $, and $ \mathbf{h}_{r}^{\textrm{NLOS}} $ are the non-line-of-sight  (NLOS) components of the corresponding channel coefficients which follow the Rayleigh fading. In addition, $ K_{1} $ is the Rician factor which is set to $ 5~\textrm{dB} $ for convenience and without loss of generality. The remaining small-scale channel coefficients are generated as circularly symmetric Gaussian random variables with zero mean and unit variance. Other parameters of the simulations are: the total transmission time period $ T = 1 $, the number of transmit antenna at the PS $ N_{T} = 6 $, number of the IRS reflecting elements $ N_{R} = 30 $, number of IoT devices $ K = 5 $, transmit power at the PS $ P_{0} = 25~\textrm{dBm} $, noise power at the AP $ \sigma^{2} = -90~\textrm{dBm} $, and energy conversion efficiency $ \eta = 0.8 $, unless otherwise specified. 

\begin{figure}[htbp]
	\centering
	\begin{minipage}[t]{0.48\textwidth}
		\centering
	\includegraphics[width=9.5cm,height = 8cm]{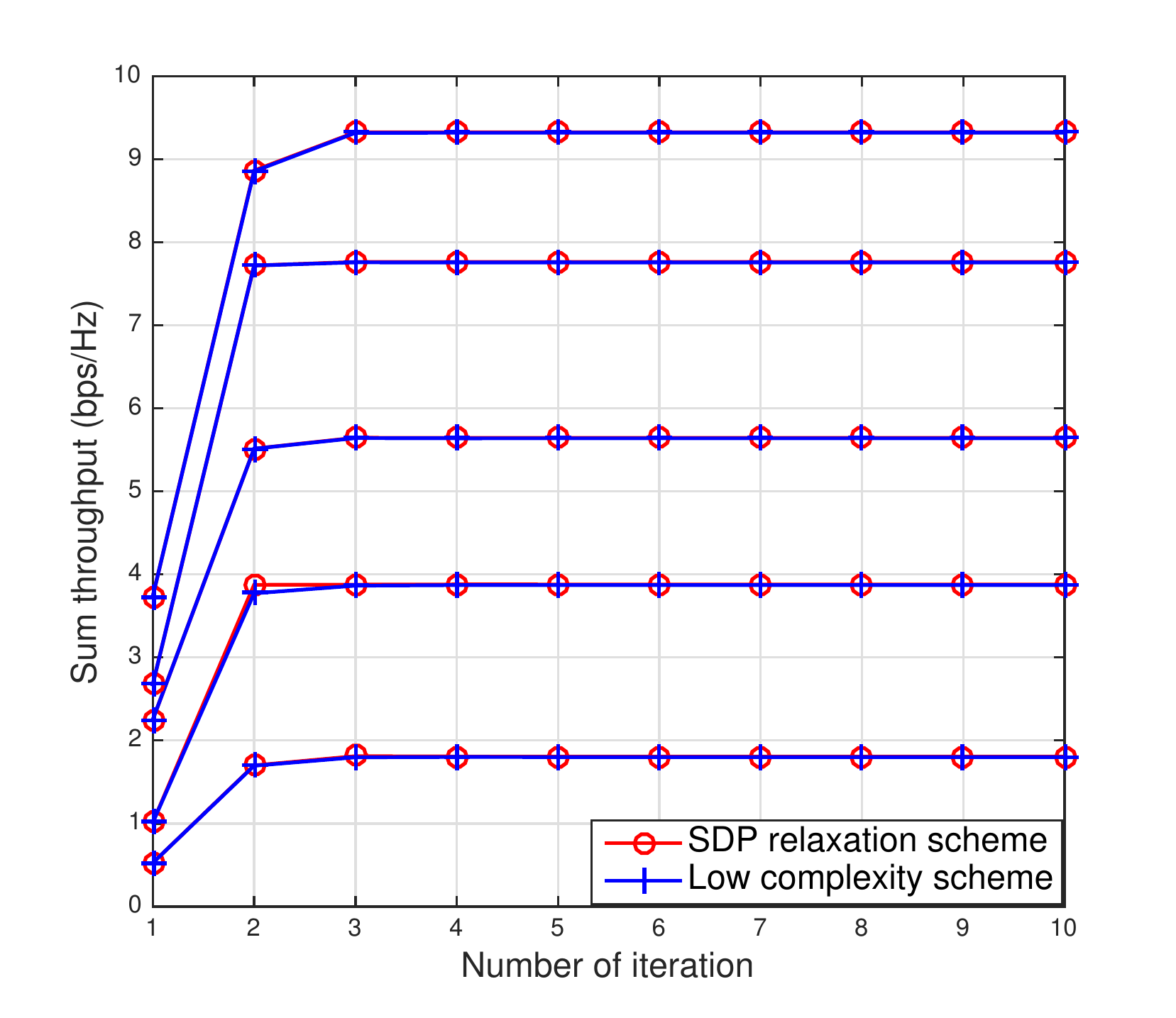}
\caption{Convergence of \emph{Algorithm} \ref{algorithm:AO1} and \emph{Algorithm} \ref{algorithm:AO2}.}
\label{fig:Iteration_plot}
	\end{minipage}
	\begin{minipage}[t]{0.48\textwidth}
		\centering
		\includegraphics[width=9.5cm,height = 8cm]{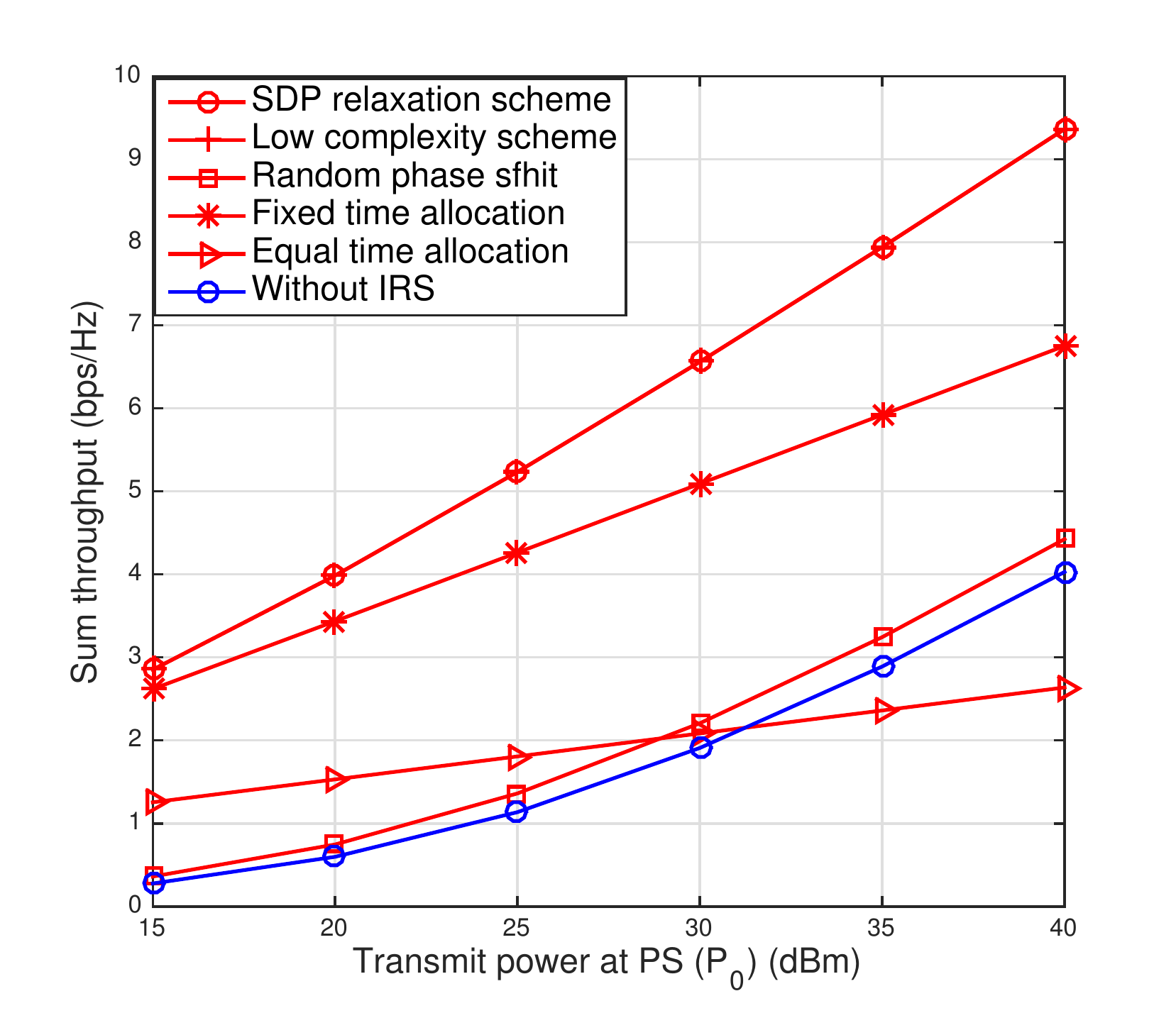}
	\caption{Sum throughput versus transmit power at the PS ($ P_{0} $).}
	\label{fig:Rate_vs_P0}
	\end{minipage}
\end{figure}

We first plot the convergence performance of \emph{Algorithm} \ref{algorithm:AO1} and \emph{Algorithm} \ref{algorithm:AO2} in Fig. \ref{fig:Iteration_plot}. From this figure, it is seen that the sum throughput increases with iterations and then achieves a converged value within five iterations, which confirms the convergence of the proposed AO algorithms. Also, the proposed low complexity scheme performance matches with the SDP relaxation scheme when the algorithms converge, which confirms the theoretical derivations as shown in Section \ref{section:Low_complexity}. 

To highlight our proposed schemes, we also evaluate the performance of the following benchmark schemes under the same parameters for comparison.
\begin{itemize}
	\item Random phase shift (RPS): the phase shifts are uniformly and randomly distributed in $ [ 0, 2\pi ) $. 
	\item Fixed time allocation (FTA) ($\tau_{0} = 0.5$). \cite{DXu_WCL_WPCN_CRN_2017}.
	\item Equal time allocation (ETA) ($ \tau_{k} = \frac{1}{K + 1} $, $ \forall k \in [0,K] $) \cite{DXu_WCL_WPCN_CRN_2017}.
	\item Without IRS \cite{RZhang_WPCN_TWC_2014}. 
\end{itemize}

In Fig. \ref{fig:Rate_vs_P0}, the sum throughput versus the transmit power at the PS $ P_{0} $ is plotted for each scheme. From this figure, one can observe that the sum throughput increases with $ P_{0} $, because more energy is collected by the IoT devices with a higher PS's transmit power which yields a larger sum throughput. In addition, the performances of the proposed SDP based AO and low complexity based AO schemes are identical and significantly outperform the benchmark schemes and the gap between the proposed schemes and the benchmark schemes becomes larger. This highlights the beneficial role of the IRS in the WPSN.
\begin{figure}[htbp]
	\centering
	\begin{minipage}[t]{0.48\textwidth}
		\centering
			\includegraphics[width=9.5cm,height = 8cm]{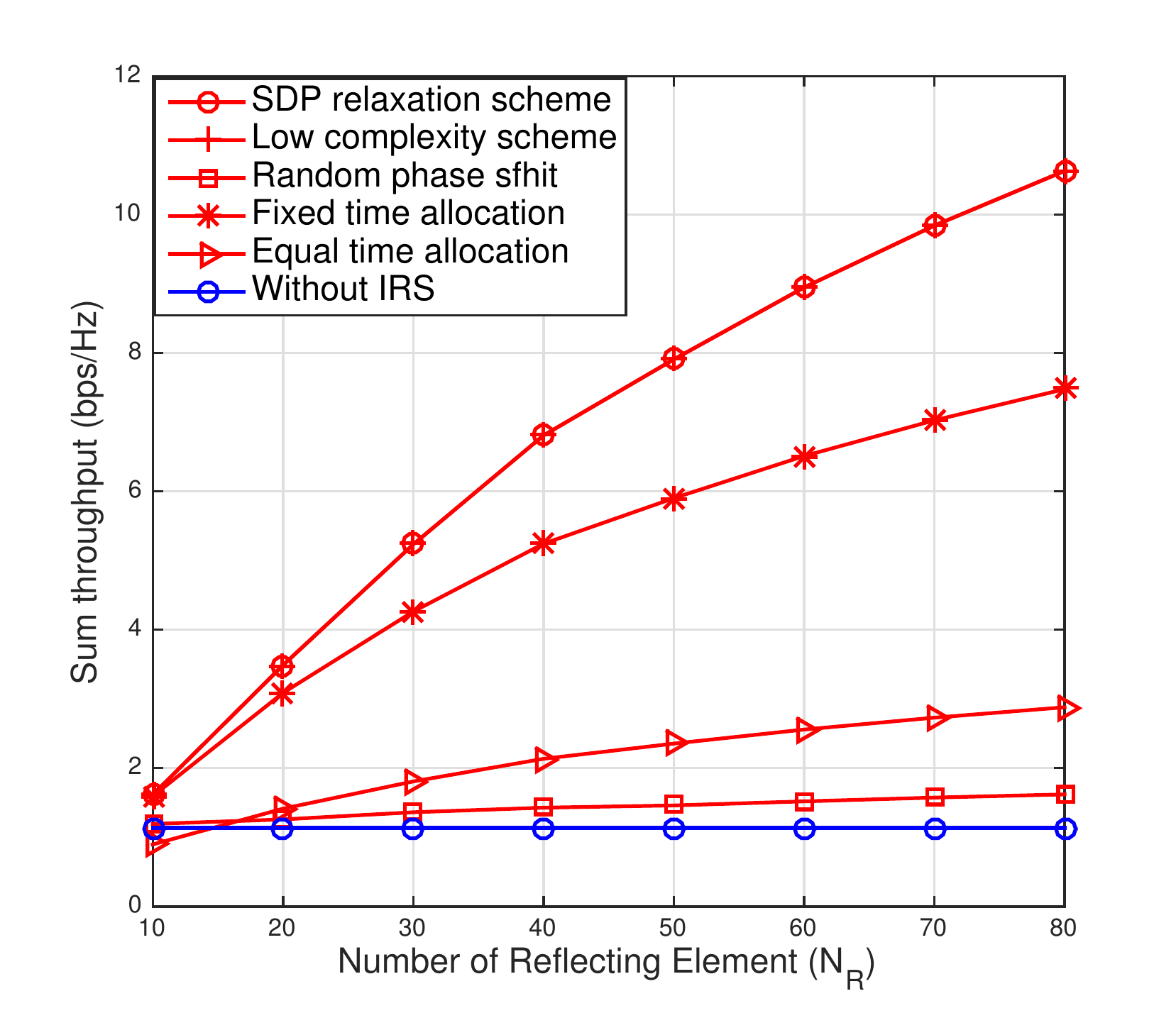}
	\caption{Sum throughput versus number of reflecting elements at the IRS $ N_{R} $.}
	\label{fig:Rate_vs_N_R}
	\end{minipage}
	\begin{minipage}[t]{0.48\textwidth}
		\centering
	\includegraphics[width=9.5cm,height = 8cm]{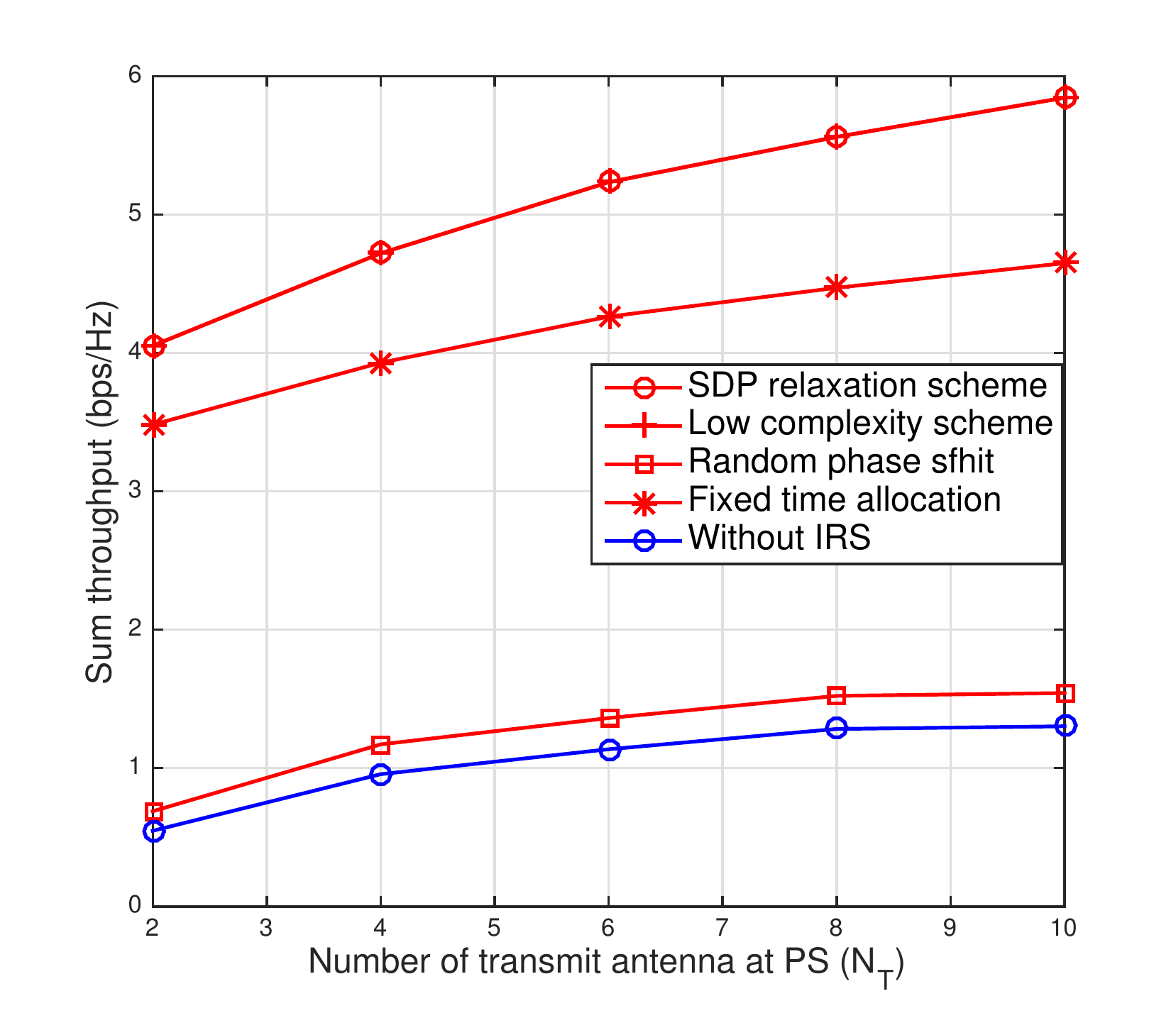}
\caption{Sum throughput versus number of transmit antenna at the PS $ N_{T} $.}
\label{fig:Rate_vs_N_T}
	\end{minipage}
\end{figure}

Next, we evaluate the sum throughput versus the number of reflecting elements at the IRS $ N_{R} $ in Fig. \ref{fig:Rate_vs_N_R}. It can observed from this figure that large-size reflecting arrays can achieve higher sum throughput than all benchmark schemes. In benchmark schemes, FTA achieves a slower increase in terms of sum throughput, since it employs a given time allocation instead of optimal time allocation, while RPS and ETA have an inconspicuous increase due to random phase shift distributed and equal time allocation. 
The scheme without IRS remains constant with $ N_{R} $ since the IRS does not involve energy and information reflections during the WET and WIT phases. 
\begin{figure}[htbp]
	\centering
	\begin{minipage}[t]{0.48\textwidth}
		\centering
		\includegraphics[width=9.5cm,height = 8cm]{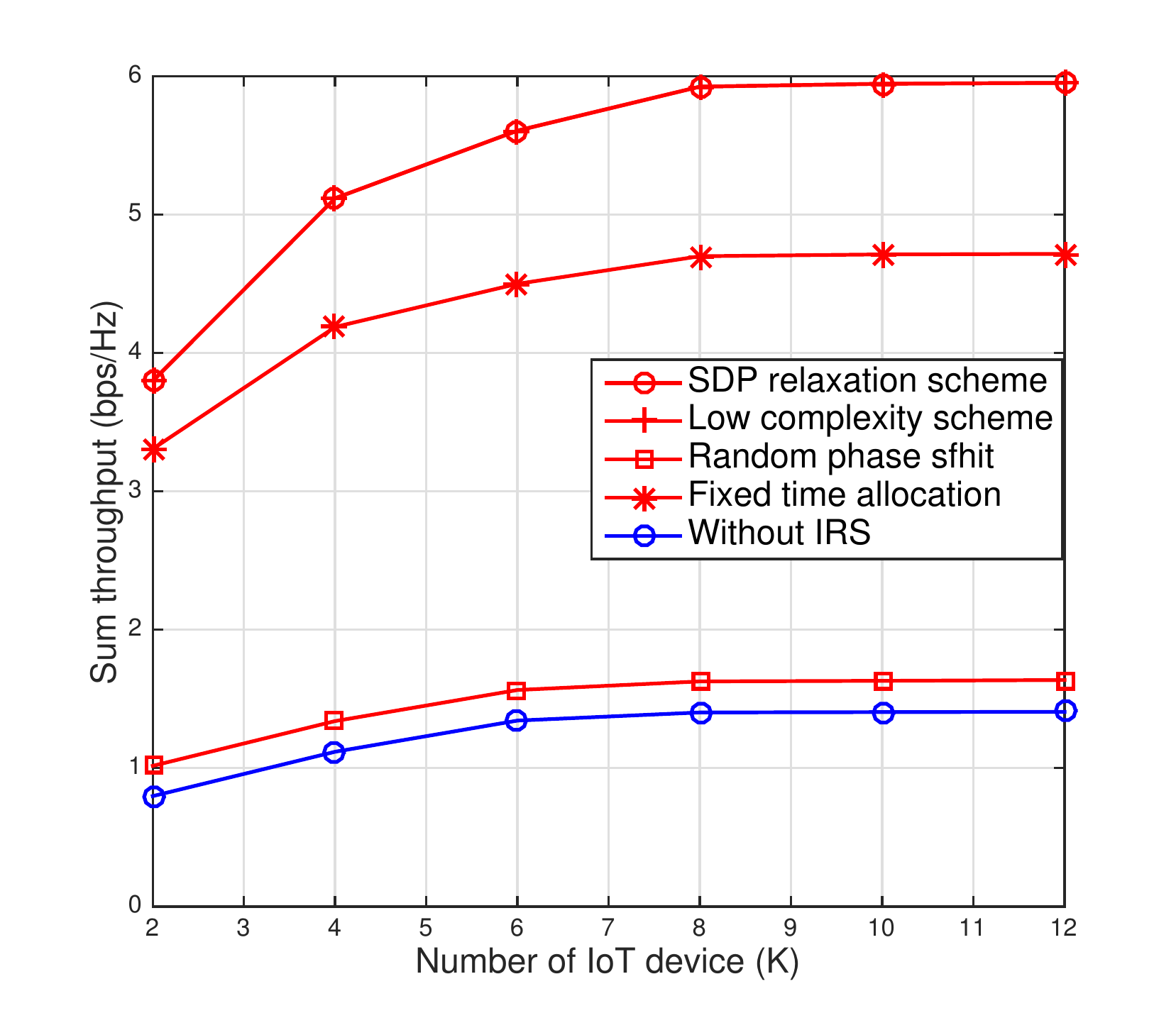}
\caption{Sum throughput versus number of IoT devices $ K$.}
\label{fig:Rate_vs_K}
	\end{minipage}
	\begin{minipage}[t]{0.48\textwidth}
		\centering
\includegraphics[width=9.5cm,height = 8cm]{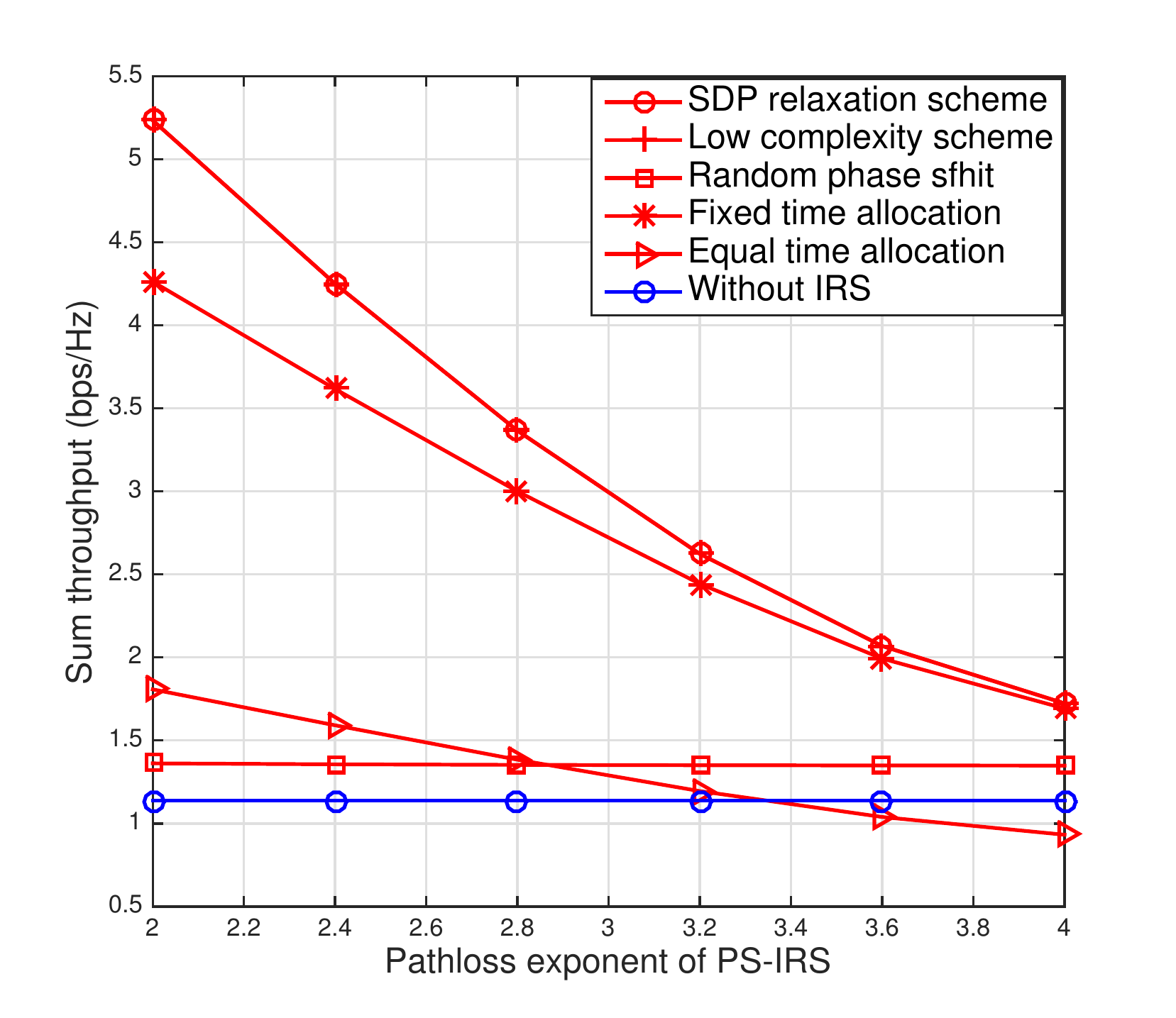}
\caption{Sum throughput versus pathloss exponent between the PS and the IRS.}
\label{fig:Rate_vs_Pathloss_PS_IRS}
	\end{minipage}
\end{figure}

Then, we demonstrate the impact of the number of transmit antennas at the PS $ N_{T} $ and IoT devices $ K $, on the sum throughput in Fig. \ref{fig:Rate_vs_N_T} and Fig. \ref{fig:Rate_vs_K}, respectively. 
Fig. \ref{fig:Rate_vs_N_T} demonstrates that the proposed schemes outperform all benchmark schemes in terms of sum throughput which increases with $ N_{T} $. This is due to the fact that a larger number of transmit antennas at the PS brings more power via energy beamforming to improve energy/information reflection efficiency. In addition, Fig. \ref{fig:Rate_vs_K} shows that a large number of IoT devices $ K $ improve the throughput performance for each scheme. To be specific, the sum throughput gradually increases with respect to $ K $ and until about $ K \geq 8 $. 
This is due to the fact that as $ K $ increases, more and more IoT devices are located farther away from the PS, the IRS, and the AP (refer to the system deployment in Fig. \ref{fig:System_deployment}) which degrades the energy and information reflection efficiencies such that the sum throughput gain becomes smaller.

\begin{figure}[htbp]
	\centering
	\begin{minipage}[t]{0.48\textwidth}
		\centering
		\includegraphics[width=9.5cm,height = 8cm]{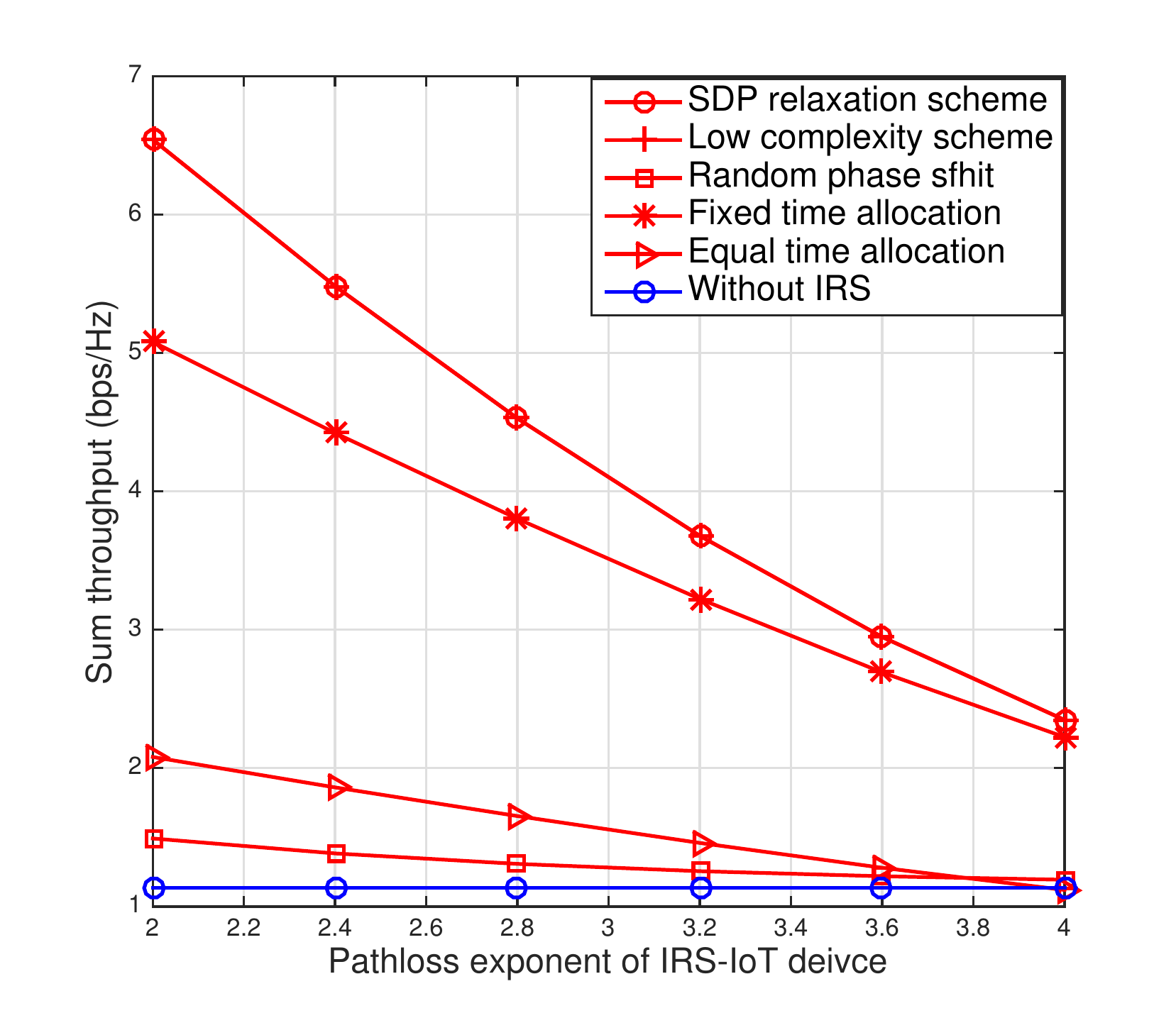}
	\caption{Sum throughput versus pathloss exponent between the IRS and the IoT devices.}
	\label{fig:Rate_vs_Pathloss_IRS_IoT}
	\end{minipage}
	\begin{minipage}[t]{0.48\textwidth}
		\centering
	\includegraphics[width=9.5cm,height = 8cm]{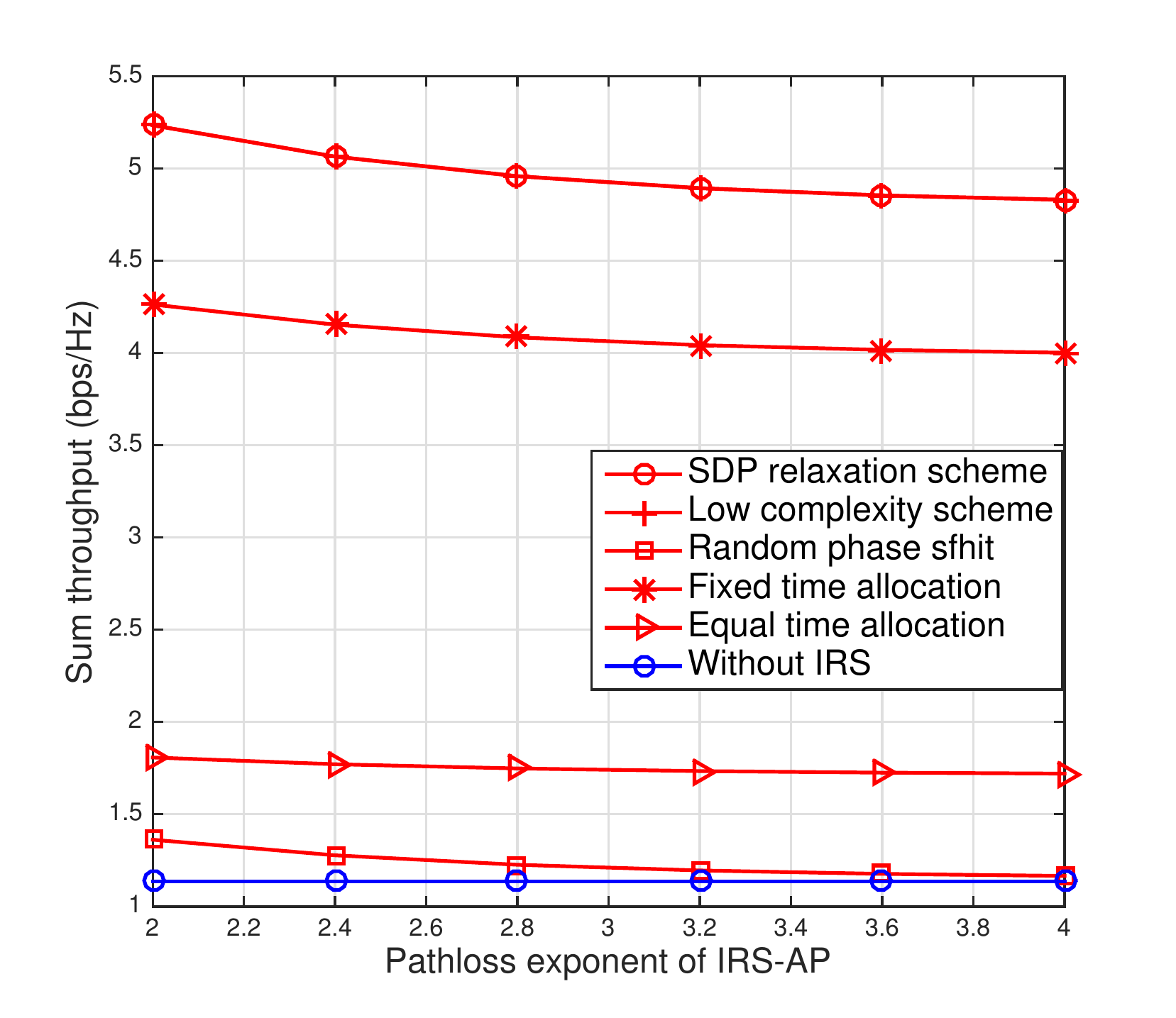}
	\caption{Sum throughput versus pathloss exponent between the IRS and the AP.}
	\label{fig:Rate_vs_Pathloss_IRS_AP}
	\end{minipage}
\end{figure}

Furthermore, the impact of the reflection's path loss model on the sum throughput is evaluated in Fig. \ref{fig:Rate_vs_Pathloss_PS_IRS} and Fig. \ref{fig:Rate_vs_Pathloss_IRS_IoT}, which investigate the energy reflection performance in the reflecting link from the PS to the IRS as well as the IRS to the IoT devices, respectively.
Fig. \ref{fig:Rate_vs_Pathloss_PS_IRS} depicts the sum throughput versus the path loss exponent between the PS and the IRS, i.e., $ \varepsilon $ is denoted by $ \varepsilon_{\textrm{PS2IRS}} $. It is apparent from this figure that the sum throughput exhibits a declining trend with $ \varepsilon_{\textrm{PS2IRS}} $ for the IRS assisted schemes, since a larger-scale fading between the PS and the IRS will result in a weaker energy signal reflected from the IRS, diminishing its benefits. Moreover, the proposed schemes have a significantly better performance than all benchmark schemes in terms of sum throughput. Specifically, the FTA can achieve a very close performance to the proposed schemes for a larger region of the path loss exponent $ \varepsilon_{\textrm{PS2IRS}} $, which demonstrates that the optimal time allocation $ \tau_{0}  $ of the proposed schemes approaches $ 0.5 $ at this point.
Since the IRS is not deployed in the traditional WPSN, its sum throughput remains constant with $ \varepsilon_{\textrm{PS2IRS}} $. 
Fig. \ref{fig:Rate_vs_Pathloss_IRS_IoT} depicts the sum throughput versus the path loss exponent between the IRS and the IoT devices, i.e., $ \varepsilon $ is denoted by $ \varepsilon_{\textrm{IRS2D}} $. Interestingly, similar trends and arguments from Fig. \ref{fig:Rate_vs_Pathloss_PS_IRS} can be also observed here in terms of the sum throughput, which has been omitted due to space limitation.
Fig. \ref{fig:Rate_vs_Pathloss_IRS_AP} depicts the sum throughput versus the path loss exponent between the IRS and the AP, i.e., $ \varepsilon $ is denoted by $ \varepsilon_{\textrm{IRS2AP}} $, which studies the information reflection of the reflecting link from the IRS to the AP. A slight throughput decreases is observed from this figure for the proposed schemes, which outperform their counterparts, i.e., RPS, FTA and ETA. This demonstrates that the information reflection of the IRS has a slight impact on the sum throughput due to large-scale fading and attenuation of the information signal transmitted by the energy harvested IoT devices. 


\begin{figure}[htbp]
	\centering
	\begin{minipage}[t]{0.48\textwidth}
		\centering
			\includegraphics[width=9.5cm,height = 8cm]{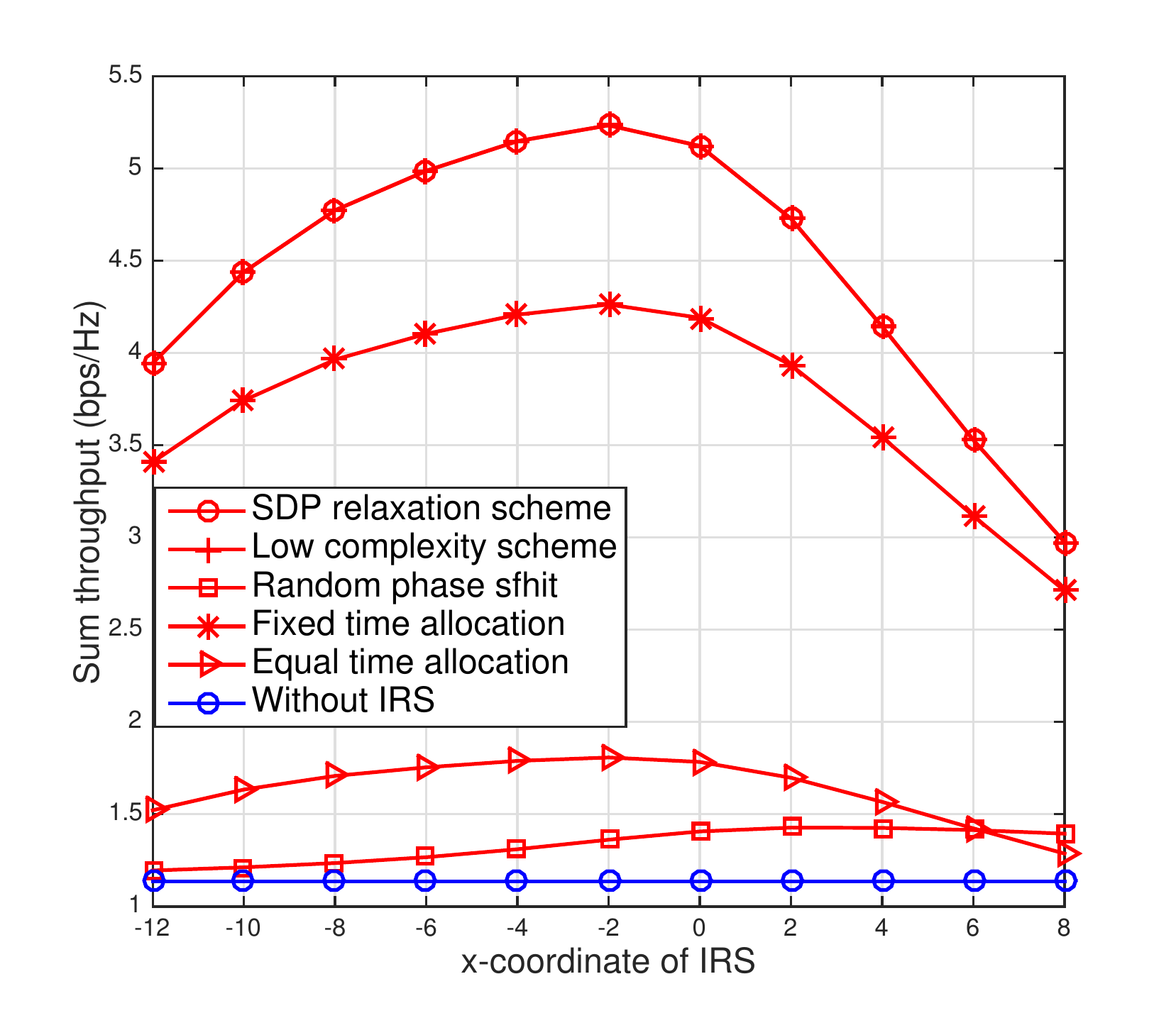}
		\caption{Sum throughput versus x-coordinate of the IRS.}
		\label{fig:Rate_vs_X_IRS}
	\end{minipage}
	\begin{minipage}[t]{0.48\textwidth}
		\centering
			\includegraphics[width=9.5cm,height = 8cm]{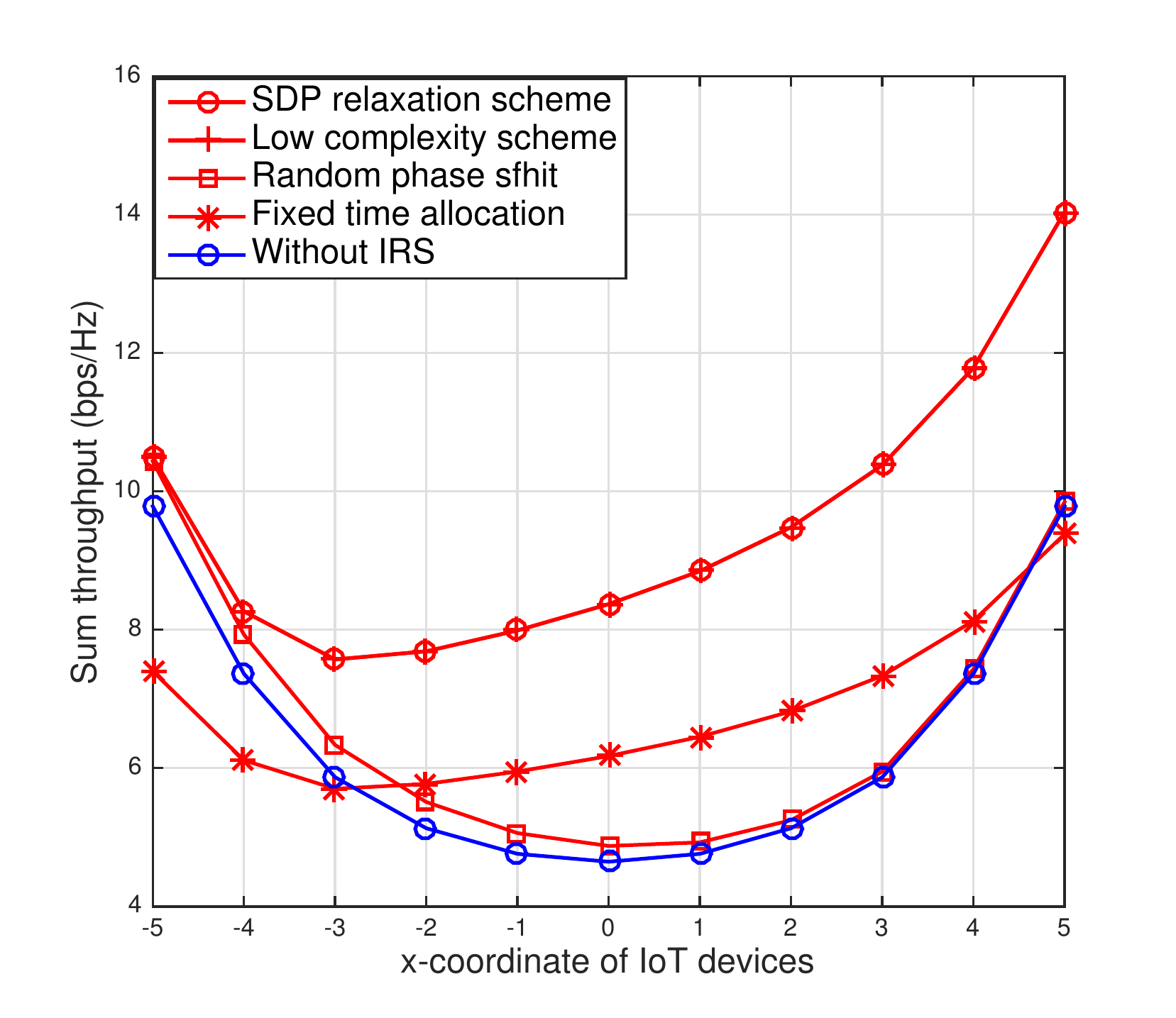}
		\caption{Sum throughput versus x-coordinate of the IoT devices.}
		\label{fig:Rate_vs_X_U5}
	\end{minipage}
\end{figure}

Finally, we evaluate the impact of the x-coordinate of the IRS (denoted by $ X_{IRS} $) and the IoT devices (denoted by $ X_{U} $) on the sum throughput in Fig. \ref{fig:Rate_vs_X_IRS} and \ref{fig:Rate_vs_X_U5}, respectively. 
In Fig. \ref{fig:Rate_vs_X_IRS}, we plot the sum throughput versus the x-coordinate of the IRS, i.e., $ X_{IRS} $. It is apparent from this figure that the sum throughput first increases and then decreases with $ X_{IRS} $. This illustrates that the optimal deployment of the IRS effectively improves the energy collection at the IoT devices so as to maximize the information reception at the AP. Fig. \ref{fig:Rate_vs_X_U5} shows the sum throughput versus x-coordinate of the IoT devices, i.e., $ X_{U} $. The PS and AP are located $ (-5,0,0) $ and $ (5,0,0) $ for convenience and without loss of generality. As seen in this figure, the sum throughput first decreases and then increases for each scheme. This illustrates that the IoT devices should be deployed either in the proximity to the AP or to the PS, resulting in a higher sum throughput.

\section{Conclusion}\label{section:Conclusion}
In this paper, we maximize the sum throughput of an IRS assisted WPSN by optimally designing the energy beamforming, the phase shifts of the WET and WIT phases and the transmission time allocation. 
We first derive the optimal phase shifts of the WIT phase in closed-form. Then, the AO algorithm was applied to alternately solve the sum throughput maximization problem. Specifically, the SDP relaxation scheme was adopted to alternately design the energy beamforming and the phase shifts of the WET phase, respectively. In addition, a low complexity scheme was proposed to derive the optimal energy beamformer, the phase shifts of WET phase and the time allocation in closed-form. Numerical results were presented to quantify the sum throughput gains of the proposed algorithm in comparison to the several benchmarks. For future work, it is interesting to investigate robust designs for the IRS assisted WPSN, where the imperfect cascaded CSIs of the PS-IRS-IoT device and IoT device-IRS-AP channels are unknown. These cascaded CSI uncertainties follow the bounded and statistical models, and the worst-case and outage probability robust beamforming designs may be adopted in the IRS assisted WPSN. However, these robust designs may demonstrate that a low complexity scheme is not feasible since it may not be possible to derive closed-form solutions of the phase shifts of the IRS, the transmission time allocation, and the energy beamforming.
\begin{appendix}
\subsection{Proof of \emph{Theorem} \ref{lemma:Theta_k_problem}}\label{appendix:Lemma_Theta_k_problem}
To prove \emph{Theorem} \ref{lemma:Theta_k_problem}, we first need to solve an equivalent sub-problem. To proceed, it is easily verified that the throughput at the AP from the $ k $-th device $ R_{k} $ is monotonically increasing with respect to $ \left | \mathbf{h}_{k} \bm{\Theta}_{k} \mathbf{h}_{r} + h_{d,k} \right |^{2} $. Thus, solving problem \eqref{eq:Original_problem_STM} with respect to $ \bm{\Theta}_{k} $ is equivalent to maximizing $ \left | \mathbf{h}_{k} \bm{\Theta}_{k} \mathbf{h}_{r} + h_{d,k} \right |^{2} $, for $ \forall k \in [1,K] $, each of which only relies on $ \bm{\Theta}_{k} $ with $ \left | \exp \left( j \alpha_{k,n} \right) \right | = 1 $ and $ \alpha_{k,n} \in [ 0, 2\pi ] $. Thus, we solve the following sub-problem with respect to $ \bm{\Theta}_{k} $, $ \forall k \in [1,K] $, instead of solving \eqref{eq:Original_problem_STM} 
\begin{align}\label{eq:Theta_k_problem}
	\max_{ \bm{\Theta}_{k} }&~  \left | \mathbf{h}_{k} \bm{\Theta}_{k} \mathbf{h}_{r} + h_{d,k} \right |^{2} \nonumber\\ 
	s.t.&~ \left | \exp \left( j \alpha_{k,n} \right)  \right | = 1, ~\forall k \in [1,K],~\forall n \in [1,N_{R}].
\end{align}
In order to solve \eqref{eq:Theta_k_problem}, its objective function can be equivalently modified as 
\begin{align} 
	| \mathbf{h}_{k} \bm{\Theta}_{k} \mathbf{h}_{r} + h_{d,k} |^{2} 
	&~= | \bm{\theta}_{k} \mathbf{b}_{k} + h_{d,k} |^{2},~\forall k \in [1,K],\label{eq:R2} 
\end{align}
where $ \mathbf{b}_{k} = \textrm{diag}(\mathbf{h}_{k}) \mathbf{h}_{r} $, $ \bm{\theta}_{k} = \left[ \theta_{k,1}, ..., \theta_{k,N_{R}} \right] = \left[ \exp(j\alpha_{k,1}),..., \exp(j\alpha_{K,N_{R}}) \right]$, $ | \bm{\theta}_{k}(n) | = 1 $, $\forall k \in [1,K],~n \in [1,N_{R}] $.
To proceed, we apply the following triangle inequality \cite{QWu_TWC_2019,SZBi_IRS_WCL_2020}
\begin{align}\label{eq:Upper_bound_Theta_k}
	\!\!\! \!\! | \bm{\theta}_{k} \mathbf{b}_{k} \!+\! h_{d,k} | \! \leq \! \sum_{n = 1}^{N} | \theta_{k,n} \mathbf{b}_{k}(n) | \!+\! | h_{d,k} | \!=\! \sum_{n = 1}^{N} | \mathbf{b}_{k}(n) | \!+\! | h_{d,k} |,
\end{align}
where $ \mathbf{b}_{k}(n) $ is the $ n $-th element of $ \mathbf{b}_{k} $ and the equality holds with  $ | \theta_{k,n}| = 1 $ for $ n \in [1,N_{R}] $. We obtain the upper bound
in \eqref{eq:Upper_bound_Theta_k} via $ \alpha_{k,n} = \arg (h_{d,k}) - \arg \left( \mathbf{b}_{k}(n) \right) $ and $ \theta_{k,n} = \exp(j \alpha_{k,n}) $, where $ \arg(\cdot) $ is the the phase operator. The optimal solution to problem \eqref{eq:Theta_k_problem} is denoted by $ \bm{\theta}_{k}^{*} $, and equivalently the optimal phase shift matrix $ \bm{\Theta}_{k}^{*} $ is found from $ \bm{\theta}_{k}^{*} $. 
	\subsection{Proof of \emph{Proposition} \ref{proposition:Theta_k}}\label{appendix:Theta_k}
	By applying a few basic mathematical manipulations, the term $ \left | \mathbf{h}_{k} \bm{\Theta}_{k} \mathbf{h}_{r} + h_{d,k} \right |^{2} $ is further equivalent to 
	\begin{align}\label{eq:Reform_Theta_k}
	&\!\!\!\!\!\!\! \left | \mathbf{h}_{k} \bm{\Theta}_{k} \mathbf{h}_{r} \!+\! h_{d,k} \right |^{2} \!=\! \left | \mathbf{h}_{k} \bm{\Theta}_{k} \mathbf{h}_{r} \right |^{2} \!+\! \left | h_{d,k} \right |^{2} \nonumber\\ &~~ \!+\! 2  \left | \mathbf{h}_{k} \bm{\Theta}_{k} \mathbf{h}_{r} \right |  \left | h_{d,k} \right | \cos \left[ \arg ( h_{d,k} ) \!-\! \arg (\mathbf{h}_{k} \bm{\Theta}_{k} \mathbf{h}_{r})  \right]. 
	\end{align}
	From \eqref{eq:Reform_Theta_k}, it is easily verified that $ \left | \mathbf{h}_{k} \bm{\Theta}_{k} \mathbf{h}_{r} + h_{d,k} \right |^{2} $ achieves its maximum value if $$ \cos \left[ \arg ( h_{d,k} ) - \arg (\mathbf{h}_{k} \bm{\Theta}_{k} \mathbf{h}_{r}) \right] = 1. $$ This means that the phases of the direct and cascaded links between the $ k $-the IoT device and the AP are identical, i.e., $ \arg ( h_{d,k} ) =  \arg (\mathbf{h}_{k} \bm{\Theta}_{k} \mathbf{h}_{r}) $, thus, \eqref{eq:Relation_both_links} holds.
	This completes the proof of \emph{Proposition} \ref{proposition:Theta_k}. 
	\subsection{Proof of \emph{Lemma} \ref{lemma:Convexity}} \label{Appedix:Convexity}
	First, the objective function \eqref{eq:Problem_reformulation2_obj} is the sum of logarithm functions, i.e., $ \tau_{k} \log \left( 1 + \frac{ \tilde{t}_{k} \textrm{Tr}\left( \mathbf{\tilde{g}}_{k} \mathbf{\tilde{g}}_{k}^{H} \mathbf{V}_{0} \right)  }{\tau_{k} } \right) $, each of which preserves its concavity with respect to $ \bm{\tau} $ and $ \mathbf{V}_{0} $ since it is the perspective function of the concave function $ \log \left( 1 + \tilde{t}_{k} \textrm{Tr}\left( \mathbf{\tilde{g}}_{k} \mathbf{\tilde{g}}_{k}^{H} \mathbf{V}_{0} \right)  \right) $. 
	In addition, all the constraints \eqref{eq:Problem_reformulation2_power} and  \eqref{eq:Original_problem_STM_time} are
	affine with $ \bm{\tau} $ and $ \mathbf{V}_{0} $. We have completed the proof of \emph{Lemma} \ref{lemma:Convexity}.
	\subsection{Proof of \emph{Lemma} \ref{lemma:Rank_one_lemma}} \label{appendix:Rank_one}
		For given $ \bm{\tau} $, we consider the following problem
	\begin{subequations}\label{eq:Problem_V0_only}
		\begin{align}
		\max_{\mathbf{V}_{0} \succeq \mathbf{0}} &~ \sum_{k=1}^{K} \tau_{k} \log \left( 1 + \frac{ \tilde{t}_{k} \textrm{Tr}\left( \mathbf{\tilde{g}}_{k} \mathbf{\tilde{g}}_{k}^{H} \mathbf{V}_{0} \right)  }{\tau_{k} } \right) \label{eq:Problem_V0_only_obj} \\
		s.t. &~ \textrm{Tr}(\mathbf{V}_{0}) \leq \tau_{0} P_{0}, 
		\end{align}
	\end{subequations}
	It is observed that \eqref{eq:Problem_V0_only_obj} is a logarithmic function which is concave but non-linear. To linearize it, we take into consideration a successive convex approximation (SCA) to transform \eqref{eq:Problem_V0_only_obj} into a series of linear programming (LP) as follows
	\begin{align}\label{eq:SCA_problem}
	\mathbf{V}_{0}^{(m+1)} = \arg \max_{ \mathbf{V}_{0} \succeq \mathbf{0} }  &~ \sum_{k=1}^{K} \frac{ \tilde{t}_{k} \textrm{Tr}\left( \mathbf{\tilde{g}}_{k} \mathbf{\tilde{g}}_{k}^{H} \mathbf{V}_{0} \right) }{ 1 + \frac{ \tilde{t}_{k} \textrm{Tr}\left( \mathbf{\tilde{g}}_{k} \mathbf{\tilde{g}}_{k}^{H} \mathbf{V}_{0}^{(m)} \right)  }{\tau_{k} }  } \nonumber\\
	s.t. &~  \textrm{Tr}(\mathbf{V}_{0}) \leq \tau_{0} P_{0}
	\end{align}
	where $ \mathbf{V}_{0}^{(m)} $ denotes the optimal solution to \eqref{eq:SCA_problem} at the $ m $-th iteration. By exploiting \cite{YWHuang_Rank_TSP_2010}, it is easily verified that $ \mathbf{V}_{0}^{(m+1)} $ can be obtained by iteratively solving \eqref{eq:SCA_problem} which yields a rank-one solution at each iteration. Thus, the optimal solution to problem \eqref{eq:Problem_reformulation2}, denoted by $ \mathbf{V}_{0}^{*} $, always returns a rank-one matrix. This completes the proof of \emph{Lemma} \ref{lemma:Rank_one_lemma}.
	\subsection{Proof of \emph{Proposition} \ref{proposition:Convergence1}} \label{appendix:Convergence1}
		In order to prove \emph{Proposition} \ref{proposition:Convergence1}, 
	the feasible solution and the objective value of problem \eqref{eq:Problem_reformulation2} are denoted by $ \left( \mathbf{\tilde{w}}, \bm{\tilde{\tau}}, \bm{\tilde{\Theta}}_{0} \right) $ and $ \tilde{f} \left( \mathbf{\tilde{w}}, \bm{\tilde{\tau}}, \bm{\tilde{\Theta}}_{0} \right) $, respectively.
	From \textbf{step} (2-b and 2-c) of \emph{Algorithm} \ref{algorithm:AO1}, there exists a feasible solution to problem \eqref{eq:Problem_reformulation3}, i.e., $ (\mathbf{\tilde{w}}^{(m+1)}, \bm{\tilde{\tau}}^{(m+1)}, \bm{\tilde{\Theta}}_{0}^{(m+1)}) $, which is also feasible to problem \eqref{eq:Problem_reformulation2}.
	As such, $ \left( \mathbf{\tilde{w}}^{(m+1)}, \bm{\tilde{\tau}}^{(m+1)}, \bm{\tilde{\Theta}}_{0}^{(m)} \right) $ in \textbf{step} (2-a) is the feasible solution to \eqref{eq:Problem_reformulation2} in the $ m $-th iteration. Thus, the following relation is easily achieved 
	\begin{align}\label{eq:Relations}
	&~ \tilde{f} \left( \mathbf{\tilde{w}}^{(m)}, \bm{\tilde{\tau}}^{(m)}, \bm{\tilde{\Theta}}_{0}^{(m)} \right) \leq \tilde{f} \left( \mathbf{\tilde{w}}^{(m+1)}, \bm{\tilde{\tau}}^{(m+1)}, \bm{\tilde{\Theta}}_{0}^{(m)}  \right) \nonumber\\ &~~~~ \leq \tilde{f} \left( \mathbf{\tilde{w}}^{(m+1)}, \bm{\tilde{\tau}}^{(m+1)}, \bm{\tilde{\Theta}}_{0}^{(m+1)}  \right),
	\end{align}
	where the first inequality holds due to the fact that $ \mathbf{w}^{(m+1)} $ and $ \bm{\tau}^{(m+1)} $ are the optimal solution to problem \eqref{eq:Problem_reformulation2} for given $ \bm{\Theta}_{0}^{(m)} $ in \textbf{step} (2-a) of \emph{Algorithm} \ref{algorithm:AO1}; The second inequality holds due to the fact that $ \bm{\Theta}_{0}^{(m+1)} $ is the sub-optimal solution to problem \eqref{eq:Problem_reformulation3} for given $ \mathbf{w}^{(m+1)} $ and $ \bm{\tau}^{(m+1)} $. 
	Thus, \eqref{eq:Relations} has a monotonically increasing trend of $ \tilde{f} \left( \mathbf{\tilde{w}}, \bm{\tilde{\tau}}, \bm{\tilde{\Theta}}_{0} \right) $. Moreover, $ \tilde{f} \left( \mathbf{\tilde{w}}, \bm{\tilde{\tau}}, \bm{\tilde{\Theta}}_{0} \right) $ is an upper-bounded due to the transmit power constraint \eqref{eq:Problem_reformulation2_power}. This completes the proof of \emph{Proposition} \ref{proposition:Convergence1}.
	\subsection{Proof of \emph{Theorem} \ref{theorem:Closedform_W}}\label{appendix:Closedform_W}
	It is easily verified that \eqref{eq:IRS_WPSN_problem_reformulation2_obj} is an logarithmic function which is increasing with respect to the term $ \textrm{Tr} \left( \mathbf{\tilde{G}} \mathbf{\tilde{G}}^{H} \mathbf{W}  \right) $, thus, the optimal solution to problem \eqref{eq:IRS_WPSN_problem_reformulation2} is equivalent to solving the following problem
	\begin{align}\label{eq:Subproblem_W}
	\max_{\mathbf{W} \succeq \mathbf{0} } &~ \textrm{Tr} \left( \mathbf{\tilde{G}}^{H} \mathbf{W}  \mathbf{\tilde{G}}  \right),~
	s.t.~ \textrm{Tr} \left( \mathbf{W} \right) \leq P_{0}.
	\end{align}
	We take into consideration the eigen-decomposition of $ \mathbf{W} $ as 
	\begin{align}\label{eq:Eigendecomposition_W}
	\mathbf{W} = \mathbf{\tilde{U}} \bm{\Delta} \mathbf{\tilde{U}}^{H},
	\end{align}
	where $ \mathbf{\tilde{U}} \in \mathbb{C}^{N_{T} \times N_{T}} $ denotes a  unitary matrix, and $ \bm{\Delta} = \textrm{diag}(\varphi_{1}, ..., \varphi_{N_{T}}) $ with $ \varphi_{1} \geq , ... , \varphi_{N_{T}} $ as well as $ \sum_{m = 1}^{N_{T}} \varphi_{m} \leq P_{0} $. By substituting \eqref{eq:Eigendecomposition_W} into the objective function \eqref{eq:Subproblem_W}, we have 
	\begin{align}\label{eq:Decomposition_equality}
	\textrm{Tr} \left( \mathbf{\tilde{G}}^{H} \mathbf{W}  \mathbf{\tilde{G}}  \right) &~ = \textrm{Tr} \left(  \mathbf{\tilde{G}}^{H} \mathbf{\tilde{U}} \bm{\Delta} \mathbf{\tilde{U}}^{H} \mathbf{\tilde{G}} \right) = \textrm{Tr}\left( \mathbf{\bar{G}}^{H} \bm{\Delta} \mathbf{\bar{G}} \right) \nonumber\\ &~ = 
	\sum_{m =1}^{N_{T}} \varphi_{m} \|\mathbf{\bar{g}}_{m}\|^{2} \leq P_{0} \|\mathbf{\bar{g}}_{1}\|^{2},
	\end{align}
	where $ \mathbf{\bar{G}}^{H} =  \mathbf{\tilde{G}}^{H} \mathbf{\tilde{U}}  = \left[ \mathbf{\bar{g}}_{1}, ..., \mathbf{\bar{g}}_{N_{T}} \right] $ with $ \mathbf{\bar{g}}_{m} \in \mathbb{C}^{K \times 1} $. Note that the equality in \eqref{eq:Decomposition_equality} holds if $ \| \mathbf{\bar{g}}_{1} \|^{2} = \max_{m \in [1,N_{T}]} \| \mathbf{\bar{g}}_{m} \|^{2} $ and $ [\varphi_{1}, \varphi_{2}, ... , \varphi_{N_{T}}] = [P_{0}, 0, ... , 0] $. Furthermore, the optimal $ \bm{\Delta} $ can be derived as 
	\begin{align}\label{eq:Delta}
	\bm{\Delta} = \textrm{diag} \left( P_{0}, 0, ... , 0 \right).
	\end{align}
	
	Let $ \mathbf{\tilde{U}} = \left[\mathbf{\tilde{u}}_{1}, ... ,  \mathbf{\tilde{u}}_{N_{T}} \right] $ with $ \mathbf{\tilde{u}}_{m} \in \mathbb{C}^{N_{T} \times 1} $, we substitute \eqref{eq:Delta} into \eqref{eq:Eigendecomposition_W}, the optimal solution to $ \mathbf{W} = P_{0} \mathbf{\tilde{u}}_{1} \mathbf{\tilde{u}}_{1}^{H} $. 
	Moreover, we define $ N = \min(N_{T}, K) $ and denote the singular value decomposition (SVD) of $ \mathbf{\tilde{G}} $ as 
	\begin{align}\label{eq:SVD_G_tilde}
	\mathbf{\tilde{G}}  = \mathbf{V}_{\mathbf{\tilde{G}}} \bm{\Gamma}^{\frac{1}{2}} \mathbf{U}_{\mathbf{\tilde{G}}}^{H},
	\end{align}
	where $  \mathbf{V}_{\mathbf{\tilde{G}}} \in \mathbb{C}^{N_{T} \times N} $ and $ \mathbf{U}_{\mathbf{\tilde{G}}}^{H} \in \mathbb{C}^{N \times N_{T}} $ each of which is a unitary matrix, and $ \bm{\Gamma} = \textrm{diag}\left( \rho_{1}, ..., \rho_{N} \right) $ with $ \rho_{1} \geq , ..., \geq \rho_{N} $. Note that $ \mathbf{U}_{\mathbf{\tilde{G}}} = [\mathbf{u}_{1}, ... , \mathbf{u}_{N}] $, where $ \mathbf{u}_{i} \in \mathbb{C}^{N_{T} \times 1}, (\forall i \in [1,N]) $ denotes the $ i $-th column of $ \mathbf{U}_{\mathbf{\tilde{G}}} $.
	
	Due to the SVD in \eqref{eq:SVD_G_tilde}, $ \|\mathbf{\bar{g}}_{1}\|^{2} $ is the maximum for all terms $ \|\mathbf{\bar{g}}_{m}\|^{2} $ if and only if $ \mathbf{\tilde{u}}_{1} $ is also the first column of $ \mathbf{U}_{\mathbf{\tilde{G}}} $, i.e., $ \mathbf{\tilde{u}}_{1} $ is equivalent to $ \mathbf{u}_{1} $, 
	which is also known as the eigenvector associated with the maximum eigenvalue of $ \mathbf{\tilde{G}} \mathbf{\tilde{G}}^{H} $, denoted by $ \rho_{1} $. 
	Thus, the optimal solution to problem \eqref{eq:Subproblem_W} is given by $ \mathbf{W}^{*} =  P_{0} \bm{\nu}_{\max} \left( \mathbf{\tilde{G}} \mathbf{\tilde{G}}^{H} \right)\bm{\nu}_{\max} \left( \mathbf{\tilde{G}} \mathbf{\tilde{G}}^{H} \right)^{H} $, 
	which completes the proof of \emph{Theorem} \ref{theorem:Closedform_W}.
	\subsection{Proof of \emph{Theorem} \ref{theorem:Optimal_tau0}}\label{appendix:Optimal_tau0}
		In order to prove \emph{Theorem} \ref{eq:Optimal_tau0}, we first rewrite the objective function in \eqref{eq:Subproblem_tau0} as 
	\begin{align}
	f(\tau_{0}) = (T - \tau_{0}) \log \left( 1 + \frac{\tau_{0} P_{0} \rho_{\max} \left( \mathbf{\tilde{G}} \mathbf{\tilde{G}}^{H} \right) }{ T - \tau_{0} } \right).
	\end{align}
	We take the first-order derivative of $ f(\tau_{0}) $ with respect to $ \tau_{0} $ and set it to zero, 
	\begin{align}\label{eq:Tau0_equality}
	&~ \frac{ \partial f(\tau_{0}) }{ \partial \tau_{0} } = - \log \left( 1 + \frac{\tau_{0} P_{0} \rho_{\max} \left( \mathbf{\tilde{G}} \mathbf{\tilde{G}}^{H} \right) }{ T - \tau_{0} } \right) \nonumber\\ &~~~~~~~+ \frac{ \frac{ T  P_{0} \rho_{\max} \left( \mathbf{\tilde{G}} \mathbf{\tilde{G}}^{H} \right) }{ T - \tau_{0} }  }{ 1 +  \frac{\tau_{0} P_{0} \rho_{\max} \left( \mathbf{\tilde{G}} \mathbf{\tilde{G}}^{H} \right) }{ T - \tau_{0} }  } = 0. 
	\end{align}
	Let $ x = 1 + \frac{ \tau_{0} P_{0} \rho_{\max} \left( \mathbf{\tilde{G}} \mathbf{\tilde{G}}^{H} \right) }{ T - \tau_{0} } $, \eqref{eq:Tau0_equality} is equivalently written as 
	\begin{align}\label{eq:Tau0_relation1}
	x \log (x) - x = P_{0} \rho_{\max} \left( \mathbf{\tilde{G}} \mathbf{\tilde{G}}^{H} \right) - 1.
	\end{align}
	After a series of mathematical manipulations, \eqref{eq:Tau0_relation1} is equivalent to
	\begin{align}\label{eq:Tau0_relation2}
 	\eqref{eq:Tau0_relation1} &~ \Rightarrow ~ \log \left( \frac{ x }{ \exp(1) } \right) \exp \left[ \log \left( \frac{ x }{ \exp(1) } \right) \right] \nonumber\\ &
 	= \frac{ P_{0} \rho_{\max} \left( \mathbf{\tilde{G}} \mathbf{\tilde{G}}^{H} \right) - 1 }{ \exp(1) }.
 	\end{align}
 By exploiting the Lambert $ \mathcal{W} $ function, i.e., $ \bar{x} \exp(\bar{x})= \bar{y} \Rightarrow \bar{x} = \mathcal{W}(\bar{y}) $, 
 \begin{align}\label{eq:Tau0_relation3}
 \eqref{eq:Tau0_relation2} &~ 
 x = \exp \left[  \mathcal{W} \left( \frac{ P_{0} \rho_{\max} \left( \mathbf{\tilde{G}} \mathbf{\tilde{G}}^{H} \right) - 1 }{ \exp(1) }  \right) + 1 \right].
	\end{align}
	From \eqref{eq:Tau0_relation3}, the optimal time allocation can be easily obtained in \eqref{eq:Optimal_tau0}, which completes the proof of \emph{Theorem} \ref{theorem:Optimal_tau0}. 
\end{appendix}
\bibliographystyle{ieeetr}
\bibliography{my_references}

\end{document}